\documentclass[fleqn,usenatbib]{mnras}

\usepackage{newtxtext,newtxmath}

\usepackage[T1]{fontenc}
\usepackage{ae,aecompl}

\usepackage{graphicx}	
\usepackage{amsmath}	
\usepackage{amssymb}	

\newcommand{\cloudy}{{\tt CLOUDY}}
\newcommand{\Prob}{P}
\newcommand{\prob}{P}
\newcommand{\DM}{{\rm DM}}
\newcommand{\DMff}{{\rm DM}_{\rm ff}}
\newcommand{\tff}{t_{\rm ff}}
\newcommand{\ta}{t_{\rm a}}
\newcommand{\DMffi}{{\rm DM}_{{\rm ff},i}}
\newcommand{\tffi}{t_{{\rm ff},i}}

\newcommand{\be}{\begin{equation}}
\newcommand{\ee}{\end{equation}}

\title[Unveiling the Engines of FRBs, SLSNe, and GRBs]{Unveiling the Engines of Fast Radio Bursts, Super-Luminous Supernovae, and Gamma-Ray Bursts}

\author[B.~Margalit {\it et al.}]{
Ben Margalit$^{1,}$\thanks{E-mail: \href{mailto:btm2134@columbia.edu}{btm2134@columbia.edu}},
Brian D.~Metzger$^{1}$,
Edo Berger$^{2}$,
Matt Nicholl$^{2}$,
\newauthor
Tarraneh Eftekhari$^{2}$
and
Raffaella Margutti$^{3}$
\\
$^{1}$Department of Physics and Columbia Astrophysics Laboratory, Columbia University, New York, NY 10027, USA
\\
$^{2}$Harvard-Smithsonian Center for Astrophysics, 60 Garden Street, Cambridge, MA 02138, USA
\\
$^{3}$Center for Interdisciplinary Exploration and Research in Astrophysics (CIERA) and Department of Physics and Astronomy,
\\
Northwestern University, Evanston, IL 60208, USA
}

\date{Accepted XXX. Received YYY; in original form ZZZ}

\pubyear{2017}

\begin{document}
\label{firstpage}
\pagerange{\pageref{firstpage}--\pageref{lastpage}}
\maketitle

\begin{abstract}
Young, rapidly spinning magnetars are invoked as central engines behind a diverse set of transient astrophysical phenomena, including gamma-ray bursts (GRB), super-luminous supernovae (SLSNe), fast radio bursts (FRB), and binary neutron star (NS) mergers.  However, a barrier to direct confirmation of the magnetar hypothesis is the challenge of directly observing non-thermal emission from the central engine at early times (when it is most powerful and thus detectable) due to the dense surrounding ejecta.  We present \cloudy~calculations of the time-dependent evolution of the temperature and ionization structure of expanding supernova or merger ejecta due to photo-ionization by a magnetar
engine, in order to study the escape of X-rays (absorbed by neutral gas) and radio waves (absorbed by ionized gas), as well as to assess the evolution of the local dispersion measure due to photo-ionization. 
We find that ionization breakout does not occur if the engine's ionizing luminosity decays rapidly, and that
X-rays typically escape the oxygen-rich ejecta of SLSNe only on $\sim 100 \, {\rm yr}$ timescales, consistent with current X-ray non-detections.  
We apply these results to constrain engine-driven models for the binary NS merger GW170817 and the luminous transient ASASSN-15lh.
In terms of radio transparency and dispersion measure constraints,
the repeating FRB~121102 is consistent with originating from a young, $\gtrsim 30-100 \, {\rm yr}$, magnetar similar to those inferred to power SLSNe.
We further show that its high rotation measure can be produced within the same nebula that is proposed to power the quiescent radio source observed co-located with FRB~121102.
Our results strengthen previous work suggesting that at least some FRBs may be produced by young magnetars, and motivate further study of engine powered transients.
\end{abstract}




\section{Introduction}

Neutron stars with exceptionally strong magnetic fields (``magnetars''; \citealt{Duncan&Thompson92}) are promising engines for astrophysical transients across a range of timescales and wavelengths.  The magnetized relativistic winds from young magnetars,
which are born rapidly spinning following core collapse supernovae (SNe) are candidates for powering long duration gamma-ray bursts (GRB; e.g., \citealt{Usov92, Thompson+04,Metzger+11,Beniamini+17}) and super-luminous supernovae (SLSNe; e.g., \citealt{Kasen&Bildsten10, Woosley10,Dessart+12,Metzger+14,Nicholl+17d}).  Hydrogen-poor SLSNe are a rare subset of the terminal explosions of massive stars stripped of their outer hydrogen envelopes which exhibit peak optical luminosities exceeding those of other SNe by factors of $\gtrsim 10-100$ \citep{Quimby+11,Chomiuk+11,GalYam12,Inserra+13,Nicholl+14,Liu+17,DeCia+17,Lunnan+17,Quimby+18} and which occur preferentially in small and irregular low-metallicity host galaxies, with properties broadly similar to those of long GRB hosts \citep{Lunnan+15,Chen+15,Perley+16,Japelj+16,Schulze+18}.  The merger of neutron star binaries can also create massive magnetar remnants (e.g.~\citealt{Price&Rosswog06,Metzger+08, Kiuchi+15}), which are temporarily supported against gravitational collapse by their rapid rotation; such meta-stable objects could help shape the electromagnetic counterparts to these gravitational wave sources (e.g.~\citealt{Metzger+18}).  Later in their evolution, magnetars can evolve to become sources of high energy radiation powered by dissipation of their enormous reservoirs of magnetic energy, which are observed as primarily Galactic sources of transient outbursts and giant flares (\citealt{Thompson&Duncan95}; see \citealt{Kaspi&Beloborodov17} for a review).

A new window into magnetized compact objects was opened by the discovery of fast radio bursts (FRBs) --- coherent pulses of radio emission lasting a few milliseconds that occur at an all-sky rate of $10^3-10^4$ per day above 1~Jy \citep{Lorimer+07,Keane+12, Thornton+13, Spitler+14, Ravi+15, Petroff+16, Champion+16,Lawrence+17}.  FRBs are characterized by large dispersion measures DM $\approx 300-2000 \,{\rm pc\, cm^{-3}}$, well above the contribution from propagation through the Milky Way or its halo and thus implicating an extragalactic
origin.  The cosmological distance of at least one FRB was confirmed by the discovery of a repeating FRB 121102 \citep{Spitler+14,Spitler+16} and its subsequent localization \citep{Chatterjee+17} to a dwarf star-forming galaxy at a redshift of $z=0.1927$ \citep{tbc+.2017}.  The unusual host galaxy properties are similar to those of long GRBs and SLSNe  \citep*{Metzger+17}, supporting a possible connection between FRBs and young magnetars \citep{Popov&Postnov13,Lyubarsky14,Kulkarni+14,Katz16,Lu&Kumar16,Metzger+17,Nicholl+17c,Kumar+17,Lu&Kumar17}. 

One mechanism by which a young magnetar could power a burst of coherent radio emission is through the synchrotron maser instability in the plasma behind magnetized shocks \citep{Gallant+92,Lyubarsky14}.  Such shocks could be produced by transient ejections from the magnetar which collide with the external medium at ultra-relativistic speeds.  This medium could represent the baryon-rich wind of material accumulated from the succession of previous recent flares  \citep{Beloborodov17} or, on larger scales, with the hot nebula of magnetic fields and particles confined behind the expanding SN ejecta \citep{Lyubarsky14}.  Other FRB emission mechanisms have been proposed that occur closer to the magnetar surface, such as antenna curvature emission within the magnetosphere (e.g.~\citealt{Kumar+17,Lu&Kumar17}).  Radio bursts from FRB~121102 have now been observed intermittently for over four years, with separations between bursts as short as seconds \citep{Spitler+16,Michilli+18}.  Any magnetar responsible for this behavior must be significantly more active than the Galactic population, which are largely dormant \citep{Kaspi&Beloborodov17}.  

Radio interferometric localization of FRB~1211012 \citep{mph+.2017} revealed a luminous ($\nu L_{\nu} \approx 10^{39}$~erg\,s$^{-1}$) steady radio synchrotron source coincident to within $\lesssim 0.8$ pc of the FRB location \citep{tbc+.2017}.  This could be interpreted as a nascent ``nebula'' surrounding the magnetar, powered by its rotational \citep{Metzger+17,Kashiyama&Murase17,Omand+18} or magnetic energy \citep{Beloborodov17}.   A plasma-dense environment surrounding FRB sources is supported also by the observed scattering tails following some FRB pulses \citep{Thornton+13,Ravi+15,Luan&Goldreich14} and possible evidence for plasma lensing of the bursts by intervening screens of dense ionized material (\citealt{Pen&Connor15,Cordes+17}; though much of the latter could be the ISM of the host galaxy).  

Constraints can be placed on the age, $t_{\rm age}$, of the putative compact remnant responsible for FRB~1211012 \citep{Metzger+17}.  Upper limits on the size of the quiescent radio source relative to predictions for an expanding nebula place a rough upper limit of $t_{\rm age} \lesssim 100$ yr.  On the other hand, a lower limit of $t_{\rm age} \gtrsim 20-30$ yr follows from the requirement that the supernova ejecta not attenuate the FRB radiation via free-free absorption or overproduce the observed DM or its time derivative \citep{Connor+16,Piro16,Metzger+17,Bietenholz&Bartel17}.  The young inferred age may be connected to the repeater's high activity as compared to Galactic magnetars \citep{Beloborodov17}, which are typically much older, $t_{\rm age} \sim 10^{4}$ yr.  \citet{Lu&Kumar17}, \citet{Nicholl+17}, and \citet{Law+17} show that if all FRB sources repeat in a manner similar to FRB~121102, then the birth rate of FRB-producing magnetars is consistent with those of SLSNe and long GRBs, and thus potentially with the subpopulation of magnetars born with particularly short rotation periods (high rotational energies).

Also supporting the existence of a dense electron-ion plasma surrounding FRB~121102 is its large rotation measure, RM $\sim 10^{5}$ rad m$^{-2}$ (\citealt{Michilli+18}; see also \citealt{Masui+15}).  This RM value exceeds those of other known astrophysical sources, with the exception of the flaring magnetar SGR J1745-2900 located in the Galactic Center at a projected offset of only 0.1 pc from Sgr A* \citep{Eatough+13}.  The magnetic field of the medium responsible for FRB~121102's RM exceeds\footnote{This minimum average magnetic field strength is derived under the conservative assumption that the RM-producing medium also contributes all of the DM, once the Milky Way and intergalactic medium values have been subtracted off.} $\sim 1$ mG \citep{Michilli+18}.  Though too high for the ISM of the host galaxy, the large field strength could instead be reasonably attributed to the same quiescent synchrotron nebula which is co-located with the bursting source.  The RM was furthermore observed to decline by $\sim$10\% over a 7 month interval \citep{Michilli+18}.  This may suggest that a turbulent magnetized environment surrounds the burst, as in the Galactic Center.  Alternatively, the decline may implicate secular evolution originating from the source being embedded in an expanding, dilluting magnetized medium, either from the supernova shock wave interacting with circumstellar gas (\citealt{Piro&Gaensler18}) or the burst-powered synchrotron nebula (see $\S\ref{sec:RM}$).  

Despite the growing circumstantial evidence tying young magnetars to a  range of astrophysical transients (GRB, SLSNe, FRB, NS mergers), definitive proof for this connection remains elusive and alternative models remain viable.  Long GRBs can be powered by fall-back accretion 
onto a black hole
of the ejecta of a massive star following a failed explosion \citep{Woosley93,MacFadyen&Woosley99}.  SLSNe could instead be powered through circumstellar interaction \citep{Chevalier&Irwin11,Moriya+13} or by fall-back accretion from a radially-extended star \citep{Quataert&Kasen12,Dexter&Kasen13}.  The high RM of FRB 121102 could indicate the bursting source just happens to be located in a magnetized galactic center environment close to an accreting massive black hole (e.g.~\citealt{Eatough+13}) or that such a location is somehow essential to the emission process \citep{Zhang17,Zhang18}, rather than originating from the birth nebula of a young stellar-mass compact object. 

One challenge in testing the magnetar model is our inability to directly view the central engine at early times due to the large absorbing column of the supernova or merger ejecta.  \citet{Metzger+14} propose to search for the emergence of UV or X-ray radiation from the young magnetar nebula on timescales of years after explosion, once the ejecta become transparent to bound-free absorption (see also \citealt{Kotera+13,Murase+15}).  Such transparency can occur gradually as the ejecta column density dilutes, or abruptly due to a sudden drop in opacity when a photo-ionization front driven by the nebula radiation reaches the ejecta surface \citep{Metzger+14}.  \citet{Margutti+17b} invoked such an ``ionization break-out" to explain the unusual UV re-brightening of the highly luminous optical transient ASASSN-15lh \citep{Dong+16} observed a few months following the explosion; this explanation could in principle apply regardless of whether the event was a true SLSN (e.g.~ejecta ionization by a central magnetar engine) or a tidal disruption event (stellar debris ionization by an accreting supermassive black hole; \citealt{Leloudas+16,Kruhler+18}).  While most SLSNe only show upper limits on their X-ray luminosity $L_{\rm X}$ \citep{Margutti+17}, SCP06F6 \citep{Barbary+09} was detected with $L_{\rm X} \sim 10^{44}-10^{45}$ erg~s$^{-1}$ roughly 70 days after the explosion \citep{Levan+13}.  The slowly-evolving SLSN PTF12dam \citep{Nicholl+13} also showed detectable X-ray emission 
with
$L_{\rm X} \approx 2\times 10^{40}$ erg~s$^{-1}$ (\citealt{Margutti+17}), though this could originate from star formation in the host galaxy.  

\citet{Kashiyama+16}, \citet{Omand+18} proposed to search for the emergence of late-time radio synchrotron emission from the engine-powered nebula, once the ejecta becomes transparent to free-free and synchrotron self-absorption (a similar condition as that needed for FRB emission to escape).  Searches for long-lived radio emission from magnetars have already been conducted for short GRBs,
leading to non-detections
\citep{Metzger&Bower14,Horesh+16,Fong+16}.
\citet{Metzger+17} proposed to search for FRB emission from $\gtrsim$ decade-old old SLSNe and long GRBs to directly connect these sources to the birth of young magnetars.  \citet{Nicholl+17} proposed the same idea for the magnetar remnants of binary neutron star mergers, a search conducted
following
the recent LIGO-discovered merger GW170817 (\citealt{Andreoni+17})
which yielded non-detections.  

Given the growing sample of FRBs with detections or upper limits on local contributions to their DM and RM (and in the case of repeating sources like FRB~121102, also of their time derivatives), as well as GRBs, SLSNe, and NS mergers with late-time X-ray and radio observations, 
it is essential
to revisit predictions for the time-evolving properties of the supernova/merger ejecta and the observability of the flaring magnetar or magnetar-powered nebula inside.  The ionization state of the ejecta, which is controlled by photo-ionization from the UV/X-ray flux of the central nebula \citep{Metzger+14,Metzger+17} or by passage of the reverse shock as the ejecta interacts with the ISM \citep{Piro16,Piro&Gaensler18}, controls the escape of X-ray/radio emission from the central sources and determines the local contribution to DM and $d$DM$/dt$.  Likewise, the magnetized nebula fed by the accumulation of past magnetar flares, provides both a steady synchrotron source and the dominant source of both RM and $d$RM$/dt$.  

Despite the importance of the temperature and ionization structure of the ejecta on these observables, previous studies of the ejecta properties have been semi-analytic in approach \citep{Metzger+14,Metzger+17}.  A more accurate treatment must account self-consistently for the ionization-recombination balance for all relevant atomic states for a realistic ejecta composition, including a self-consistent solution for the temperature structure of the ejecta.  Here, we perform such a calculation using the photo-ionization code \cloudy, applied at snapshots in time after the merger, which we then use to infer the evolution of DM, as well as the bound-free and free-free optical depth.  Combining the latter with physically- or observationally-motivated models for the intrinsic nebula radiation, we are able to predict the X-ray and radio light curves for individual SLSNe.

This paper is organized as follows.  In $\S\ref{sec:ejecta}$, we discuss basic properties of the engine and ejecta and describe our numerical approach.  In $\S\ref{sec:results}$ we present our 
\cloudy~results and 
describe the ejecta ionization state; properties of X-ray break-out with application to SLSNe, NS mergers and ASASSN-15lh; and the implied radio absorption and DM evolution in the context of FRBs.
In $\S\ref{sec:radio}$ we discuss the radio properties of the nebula and origin of the high RM associated with FRB~121102. We summarize our results in \S\ref{sec:conclusions}.

\begin{figure}
\centering
\includegraphics[width=0.45\textwidth]{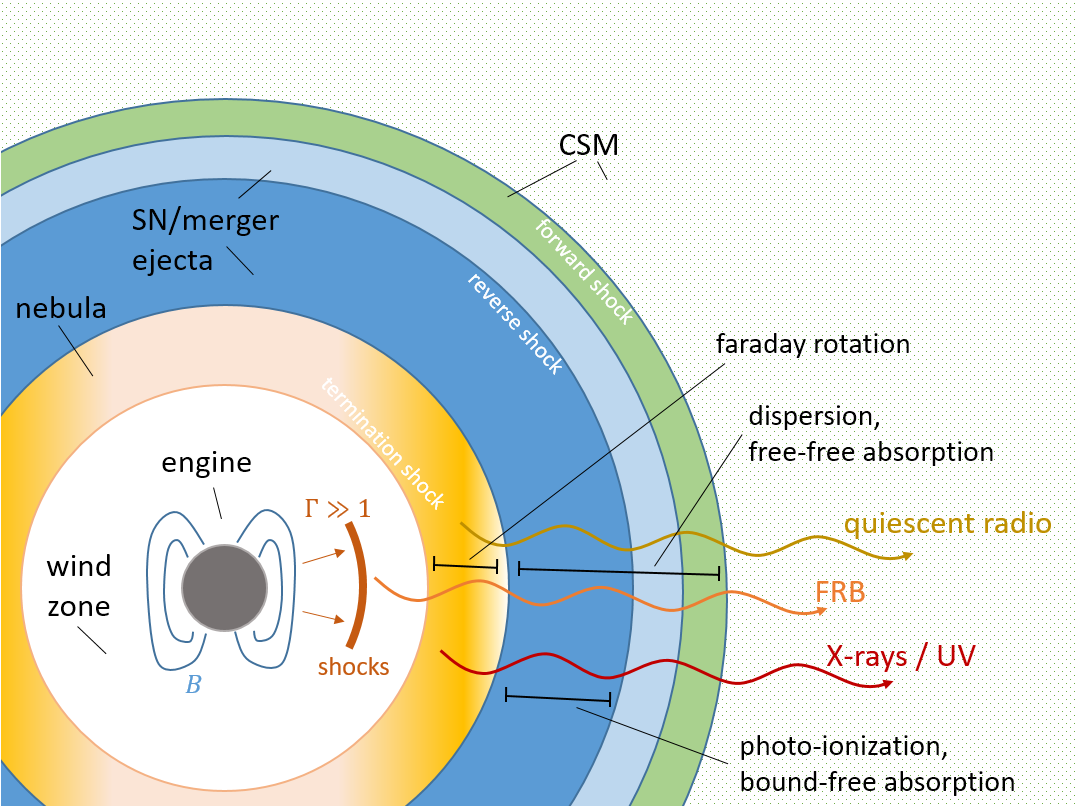}
\caption{Schematic diagram illustrating the components of the model for engine-powered transients considered in this paper. An expanding cloud of SN (or NS-merger) ejecta material (blue) envelopes a magnetar, whose spin-down and/or magnetically-driven wind shocks the ejecta interior, producing a hot nebula (yellow). UV and X-ray radiation from this nebula photo-ionizes the ejecta. An additional column of ionized material is created by the forward and reverse shocks (green and blue, respectively) generated by the ejecta's expansion into the surrounding CSM (dotted green).  FRB- or steady synchrotron radio emission produced within or interior to the nebula will undergo dispersion and free-free absorption traveling through the ejecta towards an observer, as well as scattering and Faraday rotation within the magnetized nebula.}
\label{fig:schematic}
\end{figure}

\section{Properties of the Engine and Ejecta}
\label{sec:ejecta}
\subsection{Central Engine}

A young neutron star possess two reservoirs of energy: rotational and magnetic.  Given the short spin-down timescales of magnetars, rotational energy is most important at early times following the supernova explosion or merger and is needed to power an ultra-relativistic jet in GRBs, or to inflate the nebula of radiation responsible for powering SLSNe.  
However, rotational energy cannot readily explain the large instantaneous power source of the luminous FRB emission itself, at least for FRB~121102 
and its implied age of $\gtrsim$several decades after magnetar formation
(e.g.~\citealt{Lyutikov17}).  
Magnetic energy, though generally smaller in magnitude than rotational energy, can emerge from the stellar interior more gradually over timescales of years to centuries, and thus may be responsible for powering intermittent magnetic flares responsible for FRBs, as well as an ion-electron synchrotron nebula behind the ejecta.     
\subsubsection{Rotational Energy and Ionizing Radiation}
\label{sec:rotational}

A magnetar born with an initial spin period $P_0$ and mass $M_{\rm ns} = 1.4M_{\odot}$ possesses a reservoir of rotational energy 
\begin{equation}
E_{\rm rot} \simeq
2 \times 10^{52} \, {\rm ergs} \left(\frac{P_0}{1 \, {\rm ms}}\right)^{-2}.
\label{eq:Erot}
\end{equation}
If the magnetic dipole and rotational axes are aligned, this energy is extracted through a magnetized wind on a characteristic timescale
\begin{equation}
t_{\rm rot} \simeq 4.7 \, {\rm d} \left(\frac{B}{10^{14} \, {\rm G}}\right)^{-2} \left(\frac{P_0}{1 \, {\rm ms}}\right)^2,
\label{eq:tsd}
\end{equation}
where $B$ is the surface magnetic dipole field strength.  The spin-down luminosity at time $t$ after the explosion is given by
\begin{equation}
L_{\rm rot} = \frac{E_{\rm rot} / t_{\rm rot}}{\left( 1 + t/t_{\rm rot} \right)^2}
\underset{t \gg t_{\rm rot}}\approx 8 \times 10^{40} \,{\rm ergs\,s^{-1}}\left(\frac{B}{10^{14} \, {\rm G}}\right)^{-2} \left(\frac{t}{10 \, {\rm yr}}\right)^{-2},
\label{eq:Lrot}
\end{equation}
where the last equality applies at times $t \gg t_{\rm rot}$.

Millisecond magnetars which are invoked to power cosmological GRBs must possess large magnetic fields $B \gtrsim 10^{15}$~G and short spin-down times $t_{\rm sd} \lesssim 10-10^{3}$~s, commensurate with the duration of long GRBs emission (e.g.~\citealt{Thompson+04}) or the temporally-extended X-ray emission  following some short GRBs (e.g.~\citealt{Metzger+08}).  By contrast, magnetars invoked as the power source of SLSNe must instead possess weaker magnetic fields $B \lesssim 10^{14}$~G and spin-down times of $\gtrsim$ days, comparable to the photon diffusion timescale of optical radiation through the expanding supernova ejecta (e.g.~\citealt{Kasen&Bildsten10,Woosley10,Nicholl+17d}).\footnote{However, note that the difference between the magnetar field strengths capable of powering long GRB jets versus SLSNe can be reduced if the magnetar experiences fall-back accretion because the latter enhances the spin-down luminosity of the magnetar relative to its isolated evolution \citep{Metzger+18b}.  }  

The rotationally-powered magnetar wind is ultra-relativistic with a low baryon-loading.  As in pulsar wind nebulae (PWNe), dissipation of the wind energy is expected to accelerate a power-law distribution of electrons/positrons, powering non-thermal radiation extending from radio to gamma-ray frequencies.  
The spectral energy distribution of the PWNe-like emission from a young magnetar engine is highly uncertain.  It depends on poorly understood details such as the pair multiplicity of the wind and the location of particle acceleration (e.g. close to the wind termination shock, or within regions of magnetic reconnection in the upstream wind zone; \citealt{Sironi&Spitkovsky11}).  Furthermore, while the rotationally-powered wind may be ultra-relativistic, it may be periodically interrupted or entirely subsumed by transient ejections of mildly-relativistic baryon-rich material which accompany magnetic flares (see $\S\ref{sec:magnetic}$ below) 
and may induce significant Faraday rotation.

Given these uncertainties,  we make the simplifying assumption in our photo-ionization calculations ($\S\ref{sec:results}$) that a fraction $\epsilon_i \le 1$ of the spin-down power $L_{\rm rot}(t)$ is placed into ionizing radiation, $L_{\rm e}$.  We furthermore assume a spectrum, $L_{{\rm e},\nu} \propto \nu^{-1}$ which distributes the energy equally per decade in frequency between $h \nu_{\rm min} = 1 \, {\rm eV}$ and $h \nu_{\rm max} = 100 \, {\rm keV}$.  In other words, we take
\begin{equation} \label{eq:Lnu_normalization}
\nu L_{{\rm e},\nu} = \frac{\epsilon_i L_{\rm rot}}{\log \left( \nu_{\rm max}/\nu_{\rm min} \right)}.
\end{equation}
Although this spectrum is somewhat ad hoc, a high value of $\epsilon_{i} \sim 1$ is motivated by the likelihood that the nebular electrons/positrons will be fast cooling at such a young age.  

\subsubsection{Magnetic Energy and Radio Emission}
\label{sec:magnetic}

A magnetar formed with a strong {\it interior} magnetic field of strength $B_{\star}$ contains a reservoir of magnetic energy given approximately by
\begin{equation}
E_{\rm B} \approx \frac{B_{\star}^{2}R_{\rm ns}^{3}}{6} \approx 2\times 10^{50}{\rm ergs}\,\left(\frac{B_{\star}}{5\times 10^{16} \,\,\rm G}\right)^{2},
\label{eq:EB}
\end{equation}
where $R_{\rm ns} = 12$ km is the neutron star radius.  A field strength $B_{\star} \approx 5\times 10^{16}$ G corresponds to only a few percent of equipartition with the rotational energy (equation~\ref{eq:Erot}) for $P_0 \sim 1$ ms.

If this magnetic energy emerges from the stellar interior in the form of intermittent flares, this could be responsible for powering FRB emission, for instance through coherent emission in magnetized internal shocks \citep{Lyubarsky14,Beloborodov17,Waxman17}.  The enhanced activity of FRB~121102, as compared to older Galactic magnetars, could result from more rapid leakage of the magnetic field from the neutron star interior driven by ambipolar diffusion in the core over the first $t_{\rm mag} \lesssim 10-100$ yrs following birth \citep{Beloborodov17}.  Beyond youth alone, the timescale of magnetic field diffusion $t_{\rm mag} \propto B_{\star}^{-2}$ would be shortened in magnetars for larger $B_{\star}$ \citep{BeloborodovLi16}.  Stronger interior fields might be expected if the FRB-producing sources are born rotating particularly rapidly \citep{Thompson&Duncan93}, as required for the central engines of GRBs and SLSNe.  The rate of magnetic flux leakage, and thus potentially of external flaring activity, would also be enhanced if the neutron star core cools through direct URCA reactions \citep{BeloborodovLi16}.  The latter is activated in the cores of the massive neutron stars formed from the collapse of particularly massive stars (e.g.~\citealt{Brown+18}), also implicated as the progenitors of long GRBs and SLSNe, e.g. given their observed locations in the highest star-forming regions of their host galaxies \citep{Lunnan+16,Blanchard+16}.  

In analogy with equation (\ref{eq:Lrot}) for the rotation-powered luminosity, we parameterize the time-averaged magnetic luminosity as
\be
L_{\rm mag} = \frac{E_{\rm mag}}{t_{\rm mag}}\frac{\alpha-1}{(1 + t/t_{\rm mag})^{\alpha}}. 
\label{eq:Lmag}
\ee
However, the precise value of the decay index $\alpha$ --- and indeed whether a power-law evolution is indeed even appropriate --- remains highly uncertain.  Determining this evolution with greater confidence will require future modeling of the rate of magnetic flux escape from young magnetars.

In addition to powering FRB emission itself, magnetic energy deposited in a nebula behind the ejecta could be responsible for the quiescent synchrotron radio emission \citep{Beloborodov17}.  The high rotation measure RM $\sim 10^{5}$ rad m$^{-2}$ of FRB~121102 implicates an electron-ion environment surrounding the source \citep{Michilli+18}, favoring the ion-loaded composition expected based on Galactic giant magnetar flares (e.g.~\citealt{Granot+06}), but disfavoring the relatively baryon-clean environment expected for a rotationally-powered PWN ($\S\ref{sec:rotational}$).  We estimate the RM contributed by the magnetar nebula in $\S\ref{sec:RM}$ and use it to place constraints on the required value of $E_{\rm mag}$ and baryon-loading of the magnetically-powered ejections.

\subsection{Density Profile and Composition of the Ejecta}
\label{sec:density}

We model the evolution of the supernova ejecta at radius $r$ and time $t$ as one of homologous expansion, with a broken power-law density profile
\begin{equation} \label{eq:rho_SN}
\rho(r,t) = \frac{3 M_{\rm ej}}{8 \left( v_{\rm ej} t \right)^3}
\begin{cases}
1 & \, r \leq v_{\rm ej} t
\\
\left( r / v_{\rm ej} t \right)^{-6} & \, r > v_{\rm ej} t,
\end{cases}
\end{equation}
such that the mass in material expanding above a given velocity $v = r/t$ obeys $M(>v) \propto v^{-3}$.  This particular choice for the high-velocity tail is motivated by recent numerical multi-dimensional simulations of the early-time interaction of the magnetar-inflated nebula and the surrounding ejecta (\citealt{Suzuki&Maeda17}; see also \citealt{Chen+16,Blondin&Chevalier17}).  Here $M_{\rm ej}$ is the total ejecta mass and $v_{\rm ej}$ the characteristic ejecta velocity at which the transition from a flat core to steep envelope occurs,
\begin{equation}
v_{\rm ej} = \sqrt{{10 E_{\rm ej}}/{9 M_{\rm ej}}} ,
\end{equation}
and the ejecta energy $E_{\rm ej} = E_{\rm sn} + E_{\rm rot}$ is the sum of the initial explosion energy, which we take as $E_{\rm sn} = 10^{51} \, {\rm erg}$, and the rotational energy $E_{\rm rot}(P_0)$ fed by the magnetar engine (equation~\ref{eq:Erot}).  Hence, the parameters $M_{\rm ej}$ and $P_0$ fully determine the ejecta density distribution at times greater than a few spin-down timescales (typically days for SLSNe).
Although the coupling efficiency between the magnetar spin-down luminosity and the ejecta is uncertain, the magnetar parameters we use in this study are based on fits to the photometric light-curves of SLSNe \citep{Nicholl+17d}, which only probe the energy deposited by the engine into the ejecta. It is therefore self-consistent to assume that the entirety of the observationally-inferred rotational energy is deposited within the ejecta, and eventually converted predominantly into kinetic energy (the amount of radiated energy is typically only a small fraction of the total energy).

We consider a few different assumptions about the elemental composition of the ejecta.  We first consider a spatially homogeneous composition of exclusively hydrogen, in order to whet our intuition in a simple limit and to explore ejecta ionization in related events like tidal disruption events of solar-metallicity stars.  For SLSNe and long GRBs we instead assume spatially homogeneous O-rich hydrogen-poor composition characteristic of energetic broad-lined SNe-Ic (which show similarities with SLSNe-I; e.g.~\citealt{Liu&Modjaz16,Quimby+18}).  Specifically, we adopt the composition resulting from the explosion of a $16 M_\odot$ He core model for an explosion energy $10^{52} \, {\rm erg}$ from \citet{Nakamura+01}.  The mass fraction of the first several dominant elements are: O (0.65), He (0.16), Fe (0.05), Ne (0.04), Si (0.04), Mg (0.02), and C (0.01).  

Finally, we explore a few Fe-dominated ejecta models in order to explore composition approximating the ejecta from a binary neutron star merger.  For instance, in GW170817 a large fraction of the merger ejecta was inferred to possess exclusively light $r$-process nuclei (e.g.~\citealt{Nicholl+17GW170817,Cowperthwaite+17,Villar+17}), which are expected to possess electron shell structures relatively similar to Fe \citep{Tanaka+17,Kasen+17}.

\subsection{Numerical Method to Calculate the Ejecta Ionization State} \label{sec:ionization_state}

We employ the publicly-available photo-ionization code \cloudy\footnote{http://www.nublado.org/} (version C13.1; \citealt{Ferland+13}) to calculate the ionization state of the expanding SN ejecta at different snapshots in time.  Given an incident radiation field, gas density profile (equation~\ref{eq:rho_SN}) and composition ($\S\ref{sec:density}$), \cloudy~calculates the ionization-recombination equilibrium solution and self-consistent temperature profile within the ejecta.  As discussed in $\S\ref{sec:rotational}$, we adopt as the incident radiation field from the central engine the flat spectral energy distribution normalized to a fraction of the spin-down luminosity, $L_{\rm e,\nu} \propto L_{\rm rot}(t) \nu^{-1}$ (equation~\ref{eq:Lnu_normalization}).  Though at most times of interest $t \gg t_{\rm rot}$ we are in the $L_{\rm rot} \propto t^{-2}$ portion of the decay (equation~\ref{eq:Lrot}), in our analytic discussion we  generalize the central ionizing luminosity to an arbitrary power-law decay, $L_{\rm e} \propto t^{-\alpha}$ (for instance, if $t \lesssim t_{\rm rot}$, or in case the ionizing luminosity instead tracks the release of magnetic energy; equation~\ref{eq:Lmag}).  

Since \cloudy~does not treat radiation transfer in regimes where the medium is optically-thick to electron scattering, it cannot be reliably used at very early epochs ($\lesssim 1 \, {\rm yr}$ for typical SLSNe ejecta velocities) when the Thomson optical depth through the ejecta exceeds unity.  On the other hand, our implicit assumption of ionization-recombination equilibrium is itself valid only at sufficiently early times, when the density is high enough that the recombination timescale is shorter than the heating/cooling or expansion timescales.  Assessing precisely when the equilibrium assumption breaks down is non-trivial, as it depends on the self-consistent ionization-state and temperature of the ejecta.  However, using the \cloudy~output, we estimate that ionization-recombination equilibrium holds for the dominant species of interest to the latest times of interest ($\sim$several decades), when the ejecta becomes transparent to free-free absorption at GHz frequencies (see \S\ref{sec:DM_freefree}).  

As \cloudy~is configured to calculate time-independent equilibrium solutions, we have implicitly also 
neglected adiabatic cooling in the heating balance.  This assumption holds well throughout the bulk of the ejecta initially but becomes more difficult to satisfy at late times, and is only marginally valid for the inner high-temperature ejecta layers on timescales at which the ejecta becomes transparent to free-free absorption at GHz frequencies.
Comparing the adiabatic cooling timescale ($\sim$ time since explosion) $t$ to the radiative cooling timescale within the ejecta, we find that typically $t_{\rm cool} / t \lesssim 10$ at all radii within the ejecta at the epoch when the ejecta becomes free-free transparent.  By contrast, $t_{\rm cool} / t \gtrsim 1$ only in the very inner parts of the ejecta, $r \lesssim 10^{-2} R_{\rm ej}$.  As this region contains only a small fraction of the total mass, and contributes little to the ejecta DM, free-free or bound-free optical depths, we expect that our neglect of adiabatic cooling is a reasonable approximation.  The effects of adiabatic cooling would be to moderately overestimate the temperature of the inner ejecta, and therefore slightly overestimate the DM and underestimate the free-free optical depth.


\section{Photo-ionization Results}
\label{sec:results}

\subsection{Ionization State}
\label{sec:ion}

The ionization fraction of the ejecta is defined as
\be
f_{\rm ion}(r)\equiv \frac{n_e(r)}{\sum_i Z_i n_i(r)},
\ee 
where $n_e$ is the electron number density and $Z_i$, $n_i$ are the atomic number and density, respectively, of ion $i$.
The ionization fraction along with the 
neutral fraction, $f_{\rm n}$, of the ejecta, and particularly their density-weighted averages,
\be \left\langle f_{\rm ion} \right\rangle_\rho \equiv \frac{\int f_{\rm ion} \rho dr}{\int \rho dr};\,\,\,\,\left\langle f_{\rm n} \right\rangle_\rho \equiv 1 - \left\langle f_{\rm ion}\right\rangle \label{eq:fionavg}, \ee
 are crucial ingredients in determining the X-ray ($\S\ref{sec:Xray}$) and radio ($\S\ref{sec:DM_freefree}$) opacity of the ejecta, as well as the local DM for an embedded FRB source.  This section describes the time-dependent evolution of the ionization state determined by photo-ionization, starting with pure hydrogen composition ($\S\ref{sec:H}$) and building up to the O-rich composition relevant to GRBs and SLSNe ($\S\ref{sec:O}$) and Fe-like composition relevant to NS mergers ($\S\ref{sec:Fe}$).  In $\S\ref{sec:reverse}$, we compare the DM from central photo-ionization to collisional ionization from the reverse shock traveling back through the ejecta as it begins to interact with the circumstellar medium.

\subsubsection{Pure Hydrogen Ejecta (e.g. TDE)}
\label{sec:H}
We start with the case of ejecta with a pure hydrogen composition.  This provides an illustrative example of the relevant physical processes with the added benefit of being analytically tractable.  Though not applicable to stripped-envelope supernovae, this case is relevant to photo-ionization of the hydrogen-rich stellar ejecta in a tidal disruption event.

\begin{figure}
\centering
\includegraphics[width=0.45\textwidth]{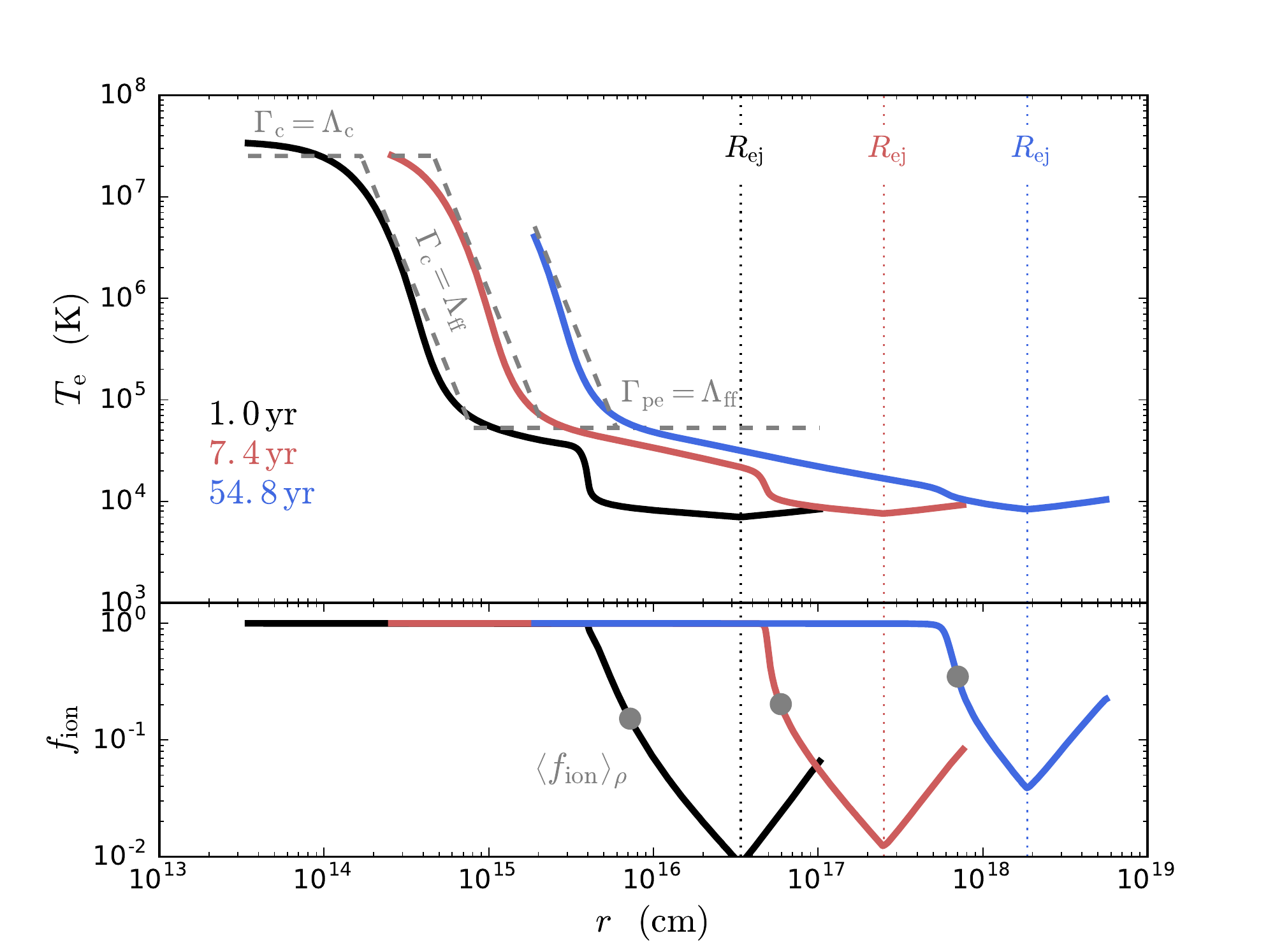}
\caption{Radial profile of ejecta ionized fraction, $f_{\rm ion}$, and electron temperature, $T_{\rm e}$, for pure hydrogen composition, shown at three successive times after the explosion as marked (black, red and blue curves, respectively).  The total ejecta mass is $M = 10 \, M_{\odot}$ and the magnetar initial spin period $P_{0} = 1$ ms and dipole field strength $B = 10^{14}$ G correspond to a characteristic ejecta velocity $v_{\rm ej} = 1.1\times 10^{4}$ km s$^{-1}$.  We assume that a fraction $\epsilon_{i} = 10^{-3}$ of the decaying spin-down luminosity of the magnetar is in ionizing radiation (equation~\ref{eq:Lnu_normalization}).  The vertical dotted lines mark the ejecta radius, $R_{\rm ej}=v_{\rm ej}t$, at which the density profile transitions from a flat inner core to a steeply-decaying envelope (equation~\ref{eq:rho_SN}). The dashed grey curves in the top panel show an analytic approximation for $T_{\rm e}$ (Appendix~\ref{sec:Appendix_Te}) applicable only within the fully-ionized region. Grey circles in the lower panel show the density-averaged ionization fraction $\left\langle f_{\rm ion} \right\rangle_\rho$ at each time. The temporal evolution of $\left\langle f_{\rm ion} \right\rangle_\rho \propto t^{1/3}$ follows the theoretical expectation (equation~\ref{eq:fion_rho_definition}) for the fiducial case in which the engine luminosity decays as $L_{\rm e} \propto t^{-2}$.} \label{fig:H_IonizationState}
\end{figure}

Figure~\ref{fig:H_IonizationState} shows time snapshots of the radial profile of the electron temperature, $T_{\rm e}(r)$, and the ionization fraction $f_{\rm ion}(r)$. The high temperature of the inner ejecta is set by a balance between Compton heating and Compton cooling, at an approximately fixed value $T_{\rm e} \sim 10^7 \, {\rm K}$ which corresponds to the ``Compton temperature" of the nebular radiation field.  At larger radii, where the radiation energy density is weaker, free-free cooling exceeds Compton, leading to a steep temperature drop, until at sufficiently large radii photo-electric heating by photo-ionization exceeds the Compton heating.  Most of the ejecta mass is concentrated at large radii, near the outer edge of the ejecta, around which $T_{\rm e}$ reaches an approximately constant value $\approx 10^{4}$ K.  The temperature profile described qualitatively above can be estimated more precisely analytically, as described in Appendix \ref{sec:Appendix_Te}, and illustrated for comparison by the dashed grey curves in Figure~\ref{fig:H_IonizationState}.

The inner portions of the ejecta are nearly fully ionized ($f_{\rm ion}\simeq1$), until reaching the ionization front at which $f_{\rm ion}$ declines to values $\lesssim 1$ (there is also an associated drop in temperature at this location).  The bottom panel of Fig.~\ref{fig:H_IonizationState} also shows the density-averaged ionization fraction, which evolves only weakly with time.  
As follows below, 
this result can be understood semi-quantitatively through a simple Stromgren sphere analysis \citep{Stromgren39}.  

Assuming that a centrally illuminating source fully ionizes a homogeneous cloud of hydrogen up to $r \lesssim R_{\rm s}$, and that $f_{\rm ion} \approx 0$ at larger radii. Equating the total production rate of ionizing photons $Q_0 = \int_{\nu_0}^\infty ( L_{\rm e,\nu} / h\nu ) d\nu \propto t^{-\alpha}$ to the total recombination rate yields the familiar Stromgren radius,
\begin{equation}
R_{\rm s} = \left(\frac{3 Q_0}{4\pi n^2 \left\langle \alpha_{\rm B} \right\rangle_m}\right)^{1/3} ,
\end{equation}
where $n$ is the density (assumed for simplicity here to be radially-constant) and $\left\langle \alpha_{\rm B} \right\rangle_m$ is the mass-averaged type-B recombination rate coefficient.
The density-averaged ionization fraction (equation~\ref{eq:fionavg}) for a spherical, homologously expanding cloud of radius $v_{\rm ej} t$ is then 
\begin{equation} \label{eq:fion_rho_definition}
\left\langle f_{\rm ion} \right\rangle_\rho \sim \frac{R_{\rm s}(t)}{v_{\rm ej} t} 
\propto M_{\rm ej}^{-2/3} v_{\rm ej} t^{1-\alpha/3} .
\end{equation}
For pure hydrogen composition, the temperature in the regions of greatest interest is regulated to an approximately constant value (see Appendix \ref{sec:Appendix_Te}), and thus $\langle \alpha_{\rm B}\rangle_m$ does not add significant temporal dependence.  Therefore $\left\langle f_{\rm ion} \right\rangle_\rho \propto t^{1/3}$ for the standard $\alpha = 2$ case where the ionizing luminosity is powered by magnetar spin-down at times $t \gg t_{\rm rot}$.

Equation~(\ref{eq:fion_rho_definition}) only applies while $R_{\rm s} < v_{\rm ej}t$, because once the ionization front reaches the outer ejecta radius the Stromgren analysis predicts $\langle f_{\rm ion}\rangle_{\rho} = 1$.  However, when calculating the bound-free X-ray opacity we are more interested in the small residual neutral fraction.  In this regime $f_{\rm n} = 1-f_{\rm ion} \ll 1$ a {\it local} version of ionization-recombination balance yields
\begin{equation}
f_{\rm n}(r,t) \approx \frac{4\pi \alpha_{\rm B} n(r,t) r^2}{\sigma_{\rm pe} Q_0(t)},
\end{equation}
where $\sigma_{\rm pe}$ is the photo-ionization cross section.  Therefore the density-averaged {\it neutral fraction} evolves as
\begin{equation}
\left\langle f_{\rm n} \right\rangle_\rho \underset{f_{\rm n} \ll 1}{\approx}
\frac{4\pi \alpha_{\rm B} M_{\rm ej}}{3 \sigma_{\rm pe} \mu m_p v_{\rm ej}t Q_0(t)}
\propto M_{\rm ej} v_{\rm ej}^{-1} t^{\alpha-1}.
\end{equation}

\subsubsection{Oxygen-rich Ejecta (SLSNe)}
\label{sec:O}
\begin{figure}
\centering
\includegraphics[width=0.45\textwidth]{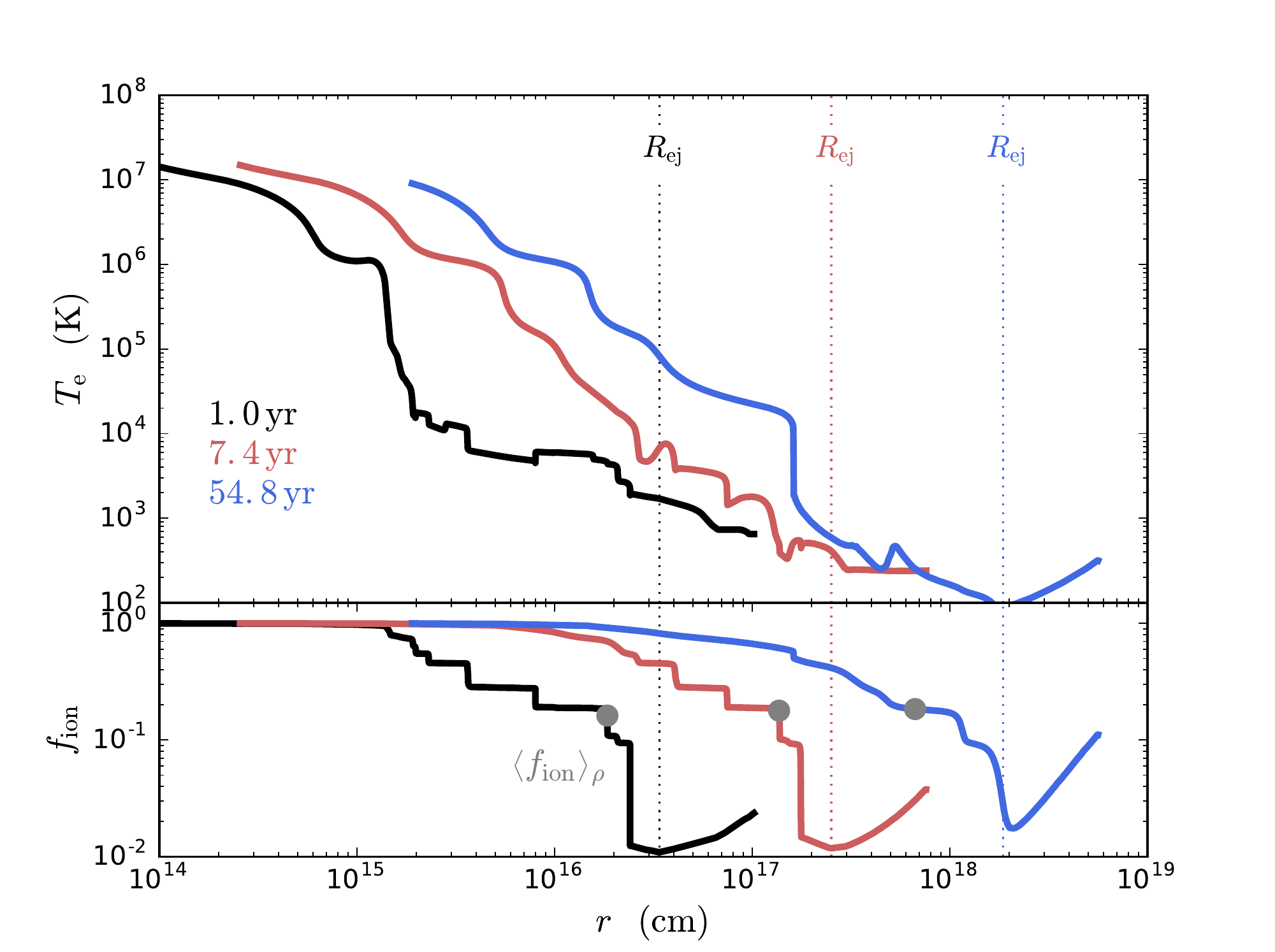}
\caption{Same as Fig.~\ref{fig:H_IonizationState}, but for the fiducial O-rich ejecta model relevant to SLSNe. The temperature and ionization profiles are significantly more complex than for the pure-hydrogen ejecta due to multiplicity of ionization fronts. The density averaged ionization fraction $\left\langle f_{\rm ion} \right\rangle_\rho$ remains roughly constant in time in this scenario.} \label{fig:O_IonizationState}
\end{figure}

The case of O-rich ejecta relevant to GRB and Type I-SLSNe cannot be simply described by the Stromgren sphere analysis due to the large number of different ionization states.
Figure \ref{fig:O_IonizationState} shows our \cloudy~calculation of the radial profile of the ionization fraction and electron temperature of the ejecta at three different epochs. In contrast to the pure hydrogen case, the radial profiles show significant structure representative of the multiple ionization fronts for different species, consistent with the picture outlined in \citet{Metzger+14,Metzger+17}.
Qualitatively, the temperature is still set by Compton heating near the inner edge of the ejecta and by photo-electric heating at large radii.  However, in the O-rich case line-cooling instead exceeds free-free cooling throughout most of the ejecta volume.

Given this analytically-untractable complexity, it is fortunate that our main interest is in global properties related to the average ionization state and temperature of the ejecta.  Empirically, for O-rich ejecta and a $L_{\rm e} \propto t^{-2}$ decaying ionizing luminosity ($\alpha=2$) we find that the density-averaged ionization fraction\footnote{However, note that one can still define an `effective' Stromgren radius $R_{\rm s, eff} \equiv \left\langle f_{\rm ion} \right\rangle_\rho v_{\rm ej}t$ (see equation~\ref{eq:fion_rho_definition})} 
remains nearly constant in time, $\left\langle f_{\rm ion} \right\rangle_\rho \propto t^0$.  
We also consider the case of a constant luminosity source, $L_{\rm e} \propto t^0$ ($\alpha=0$), for example, describing the early plateau phase of spindown evolution at $t < t_{\rm rot}$ (equation~\ref{eq:tsd}). In this case we find the density averaged ionization fraction increases as roughly $\left\langle f_{\rm ion} \right\rangle_\rho \propto t^{0.4}$ before saturating at close to complete ionization. 
As discussed further in \S\ref{sec:Xray}, these results indicate that the normally-considered picture of ionization breakout is only possible for a temporally constant or slowly-decaying ionizing radiation sources.  Stated another way, if ionization break-out does not occur by $t \lesssim t_{\rm rot}$, it will not occur at later times.

\subsubsection{Pure Iron Ejecta (NS Merger)}
\label{sec:Fe}
To explore the photo-ionization of ejecta by a long-lived central remnant in the case of binary NS mergers, we apply our methods to ejecta of mass $M_{\rm ej} = 0.05 M_\odot$ and velocity $v_{\rm ej} = 0.2$ c,  motivated by the inferred properties of the kilonova emission accompanying GW170817 \cite[e.g.][]{Cowperthwaite+17,Kasen+17,Villar+17}. Although the merger ejecta is expected to be composed of freshly-synthesized $r$-process material, the atomic data for these elements is not currently incorporated into \cloudy. For this first approach to the problem, we therefore assume an iron-rich composition, which exhibits the closest degree of complexity to at least the light $r$-process nuclei that is achievable within the current limitations.  Also note that the implicit assumption within \cloudy~that the Thompson optical depth is small does not always hold for our models at early times.  We are therefore likely underestimating the ionization fraction of the merger ejecta due to the inability of \cloudy~to treat backscattering. We continue with the calculations despite these two important caveats, leaving more accurate modeling of merger ejecta photo-ionization to future studies.

\begin{figure}
\centering
\includegraphics[width=0.45\textwidth]{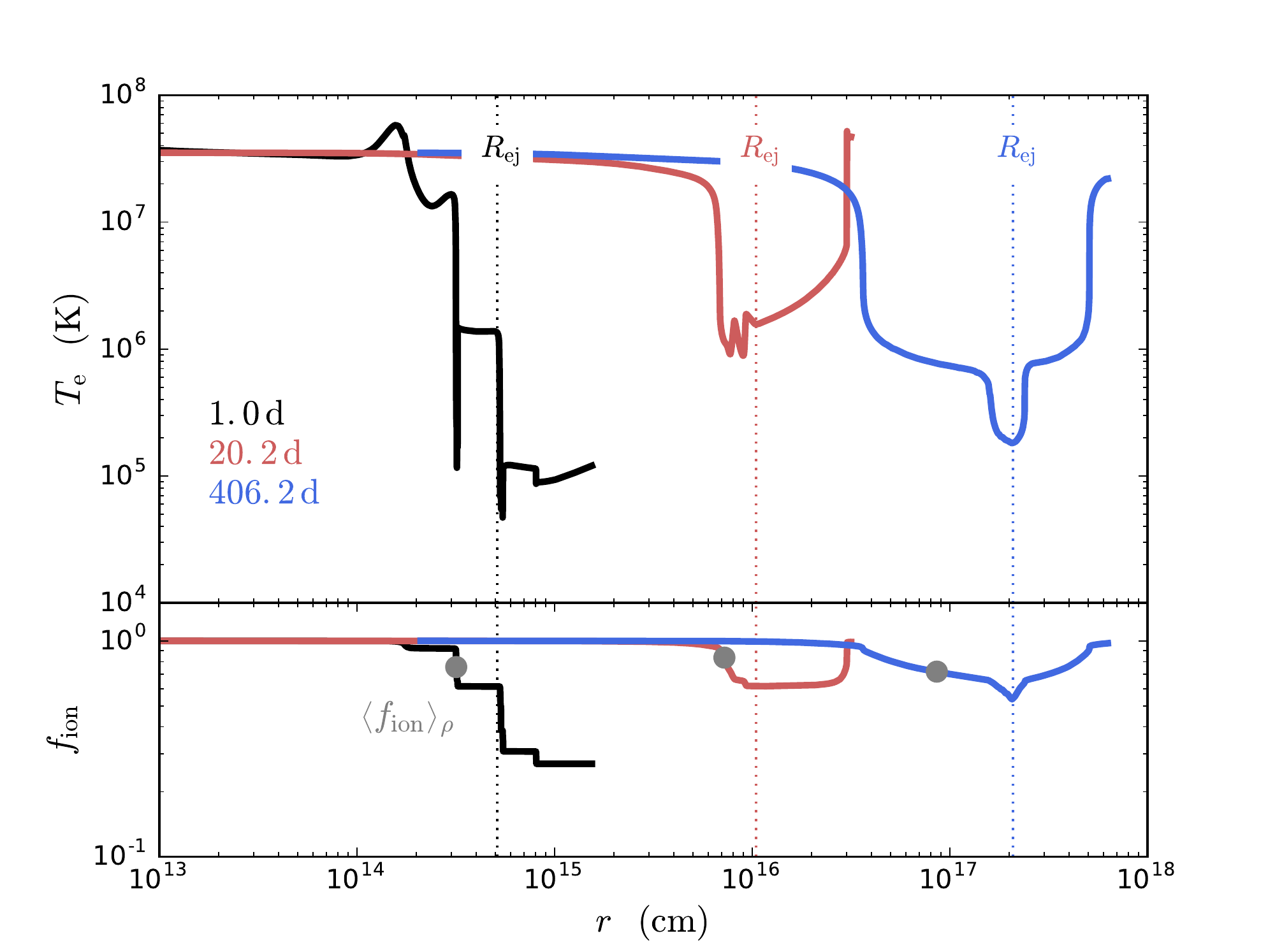}
\caption{Same as Fig.~\ref{fig:H_IonizationState}, but for pure Fe composition, meant to approximate the properties of the $r$-process ejecta from binary NS mergers.  As in the O-rich ejecta case, the value of $\left\langle f_{\rm ion} \right\rangle_\rho$ is approximate constant in time. } \label{fig:Fe_IonizationState}
\end{figure}

Figure~\ref{fig:Fe_IonizationState} shows the ionization state for the fiducial merger-ejecta model, with a  magnetar of dipole magnetic field $B = 10^{16} \, {\rm G}$ 
and initial spin-period $P_0 = 0.8 \, {\rm ms}$ (corresponding to the break-up limit for a NS, the relevant scenario given the large orbital angular momentum at merger)
as the ionizing radiation source.  The snapshots shown are earlier than in the SN case, due to the faster evolution of the merger ejecta given its lower mass and higher velocity.  Models run for lower assumed dipole fields (larger ionizing fluxes at times $t \gg t_{\rm rot}$) result in nearly complete ionization of the NS merger ejecta at all epochs.  We also find, similarly to the O-rich SLSN case, that the density-averaged ionization fraction $\left\langle f_{\rm ion} \right\rangle_\rho$ is roughly constant in time at times $t > t_{\rm rot}$, when $L_{\rm e} \propto t^{-2}$.  Again, ionization breakout appears not to be effective unless it has already taken place by $t \sim t_{\rm rot}$ (see \S\ref{sec:Xray} for further discussion).

\subsection{X-ray Light Curves}
\label{sec:Xray}

\subsubsection{Oxygen-rich Ejecta (SLSNe)}

One potential test of the magnetar model (or more, broadly, engine-powered models) for SLSNe is the onset of late-time X-ray emission, produced once ionizing radiation from the rotationally-powered nebula escapes from the expanding ejecta \citep{Metzger+14,Kotera+13}.  At early times, X-rays are attenuated by bound-free absorption in the ejecta. The X-ray optical depth through the ejecta is given by
\begin{equation} \label{eq:tau_X}
\tau_X = \int \sigma_{\rm bf} n \left( 1 - f_{\rm ion} \right) \, dr
= \frac{3 \sigma_{\rm bf} M_{\rm ej}}{8 \mu m_{\rm p} v_{\rm ej}^2 t^2}\left\langle f_{\rm n} \right\rangle_\rho ,
\end{equation}
where $n = \rho/(\mu m_p)$ and $\sigma_{\rm bf} = \int F_\nu \sigma_{\nu}(\nu) d\nu / \int F_\nu d\nu$ is the flux averaged bound-free cross section 
within the observed X-ray frequency band.

There are two ways that $\tau_{\rm X}$ can decrease below unity, initiating the X-ray light curve to rise to its peak.  First, $\tau_X$ can decrease abruptly, driven by changes in the ejecta's ionization state (i.e. because $\left\langle f_{\rm ion} \right\rangle_\rho \to 1$), a so-called `ionization breakout' \citep{Metzger+14}.  Alternatively, the condition $\tau_{\rm X} < 1$ can be achieved more gradually, due to the $\propto t^{-2}$ decrease in the ejecta column at fixed $\left\langle f_{\rm ion} \right\rangle_\rho$, a process we refer to as `expansion-dilution'.

\begin{figure}
\centering
\includegraphics[width=0.45\textwidth]{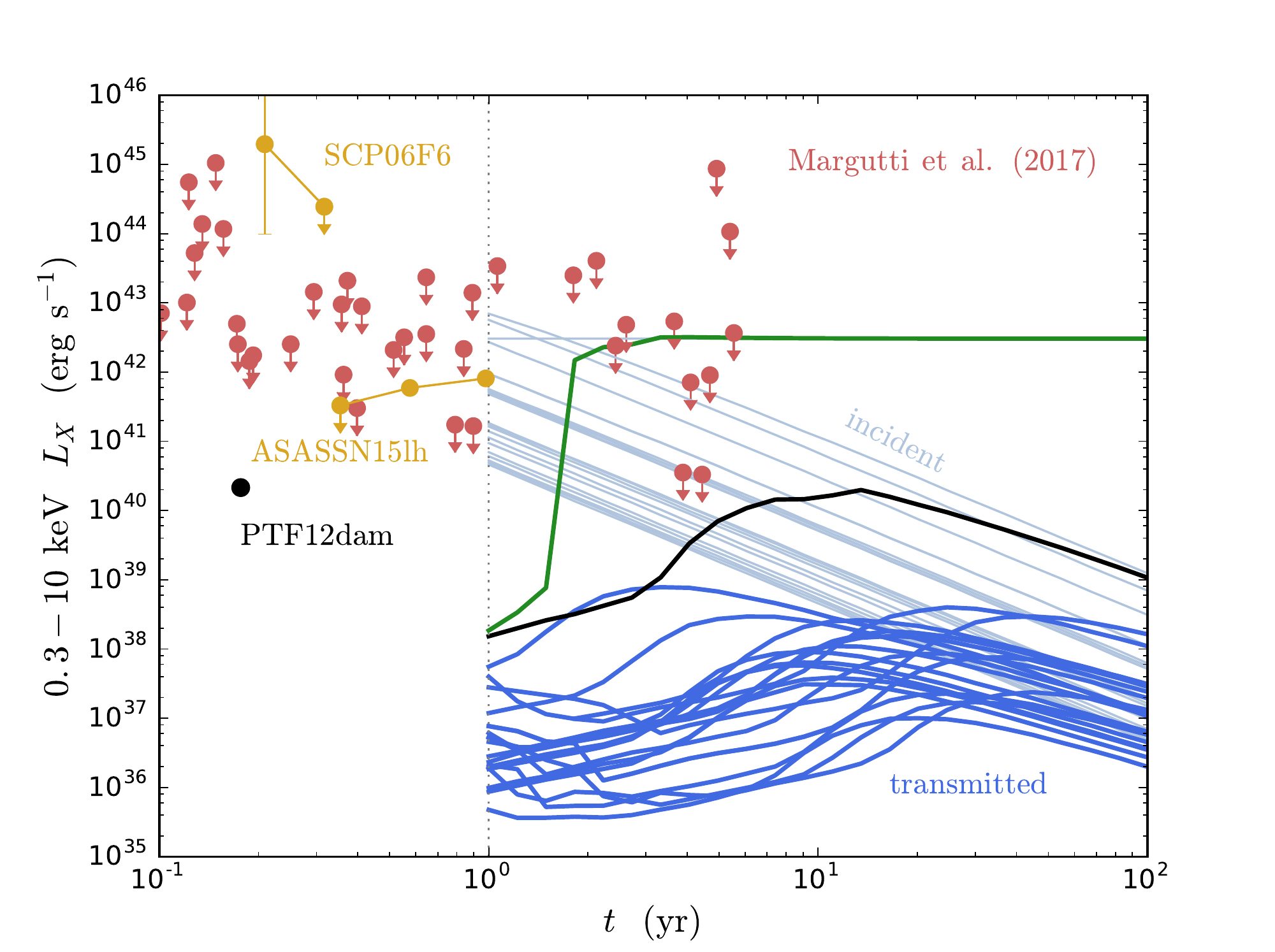}
\caption{Unattenuated X-ray luminosity from the magnetar nebula (light blue light curves) and transmitted luminosity through the supernova ejecta (dark blue light curves) from our \cloudy~calculations, for different engine and ejecta parameters from the sample of SLSNe modeled by \citet{Nicholl+17}.  We assume an efficiency $\epsilon_i = 1$ for converting spin-down luminosity into broad-band ionizing luminosity (equation~\ref{eq:Lnu_normalization}).  The nebular X-rays are initially attenuated by bound-free absorption, until the ejecta undergoes sufficient dillution for the the bound-free optical depth to decrease below unity (equation~\ref{eq:tbf}), after which time the incident and transmitted light-curves converge.   The detection of SCP06F6 \citep{Levan+13} and PTF12dam \citep{Margutti+17} are shown in yellow/black respectively, while red circles show current upper limits \citep{Margutti+17}. The X-ray light-curve of PTF12dam predicted by our fiducial model is also highlighted in black.  Despite exhibiting the strongest light-curve among SLSNe in our sample, it seems difficult to interpret the early X-ray flux from PTF12dam as originating from the central engine. A green curve shows the X-ray light-curve for an artificial model with temporally-constant ionizing luminosity, $L_{\rm rot}=10^{43} \, {\rm erg \, s}^{-1}$; this model exhibits an ionization breakout whereby X-rays escape due to a decrease in the average neutral fraction $\left\langle f_{\rm n} \right\rangle_\rho$ \citep{Metzger+14} as opposed to dillution effects.} \label{fig:x-ray}
\end{figure}

The previous section showed that $\left\langle f_{\rm ion} \right\rangle_\rho$ is approximately constant for O-rich ejecta if $L_{e} \propto t^{-2}$.   Thus, in the $t \gtrsim t_{\rm rot}$ portion of the magnetar spin-down evolution (equation~\ref{eq:Lrot}), we conclude that ionization-breakout is irrelevant and X-rays can only escape due to expansion-dilution.  Stated another way, if an ionization break-out is not achieved by $t \sim t_{\rm rot}$, then it is unlikely to be achieved at later times $t \gtrsim t_{\rm rot}$.

This behavior is apparent in Figure \ref{fig:x-ray}, which shows the transmitted X-ray light curves extracted from our \cloudy~results using parameters relevant to the sample of SLSNe in \cite{Nicholl+17} and assuming $\epsilon_{i} = 1$.  The peak luminosities, achieved on timescales of $\sim 3-30$ years, are in all cases low, $L_{\rm X} \lesssim 10^{39}$ erg s$^{-1}$ and consistent with non-detection upper-limits shown for comparison from \cite{Margutti+17}.  

If ionization-breakout were instead responsible for the X-ray escape, then the calculated light-curves would rise to peak more abruptly, when the dominant ionization front corresponding to the X-ray observing band reaches the ejecta surface.  Instead, for the ejecta-dillution scenario the light-curves evolve gradually, displaying the expected $L_X \propto (t / t_{\rm e})^{-2} \exp \left[ -(t/t_{\rm bf})^{-2} \right]$ behavior, where $t_{\rm bf}$ is the time at which $\tau_{\rm X} = 1$.  For a temporally-constant $\left\langle f_{\rm n} \right\rangle_\rho \approx 0.5 $ the latter can be estimated as (equation~\ref{eq:tau_X})
\begin{equation}
t_{\rm bf} \approx 130 \, {\rm yr} \left(\frac{M_{\rm ej}}{10 M_\odot}\right)^{1/2} \left(\frac{v_{\rm ej}}{10^4 \, {\rm km \, s}^{-1}}\right)^{-1} \left(\frac{Z}{8}\right)^{-3/2} \left( \frac{\left\langle f_{\rm n} \right\rangle_\rho}{0.5}\right)^{1/2},
\label{eq:tbf}
\end{equation}
where we have taken $\mu \approx 2Z$ and have approximated the bound-free opacity $\sigma_{\rm bf} \approx 8 \times 10^{-18} \, {\rm cm}^2 \, Z^{-2}$ by its value near the ionization threshold frequency for hydrogen-like ion of atomic number $Z$.

To explore a particularly optimistic case in which ionization break-out might occur, we calculate models with a temporally-constant ionizing radiation source. These models, presented also in \S\ref{sec:O}, exhibit a temporal increase in the mean ionization fraction which dominates over the $t^{-2}$ expansion-dilution of the ejecta, such that ionization break-out occurs. The transmitted X-ray luminosity of one such model, with an ionizing luminosity of $10^{43} \, {\rm erg \, s}^{-1}$, is depicted by the green curve in Fig.~\ref{fig:x-ray}. The sharp transition at $t \approx 2 \, {\rm yr}$ due to the rapid increase in the ejecta's ionization state marks the onset of ionization break-out. This behavior differs significantly from the slow-evolving light-curves of $L_{\rm e} \propto t^{-2}$ ionizing radiation sources (blue curves in Fig.~\ref{fig:x-ray}), which are characterized instead by ejecta-dilution.  
The qualitative result is not strongly dependent on the value of the assumed fixed ionizing luminosity source. For example, even lower luminosities of $10^{41} \, {\rm erg \, s}^{-1}$ induced a rapid increase in the ejecta's ionization state and led to an ionization break-out on timescales of $\sim 5 \, {\rm yr}$. We expect this behavior as long as the luminosity is sufficient to drive $\left\langle f_{\rm ion} \right\rangle_\rho \to 1$ faster than the expansion-dilution effect.

Finally, we note that density inhomogeneities in the ejecta, e.g. due to Rayleigh-Taylor instabilities caused by the PWN accelerating into this medium \citep{Chen+16,Suzuki&Maeda17,Blondin&Chevalier17}, can allow X-rays to escape at earlier times than predicted by our spherical models. \cite{Blondin&Chevalier17} show that the column density along certain viewpoints can decrease by more than an order of magnitude due to such inhomogeneities. This would allow radiation leakage reaching some observers at $\sim$three times earlier than predicted by the spherical models. Variability of the column density due to turbulent motions expected within this inhomogeneous ejecta may also affect the X-ray light-curves, and we leave study of such effects to further work.

\subsubsection{ASASSN-15lh} \label{sec:ASASSN15lh}

As a test case illustrating the strong dependence of the transmitted X-ray flux on model parameters, we examine the X-ray and UV emission of the very luminous transient ASASSN-15lh \citep{Dong+16}.  The nature of ASASSN-15lh has been debated extensively, the two prominent models interpreting the event as either a SLSN or a tidal disruption event \citep[e.g.][]{Metzger+15,Dong+16,Leloudas+16,Sukhbold&Woosley16,Margutti+17b,Kruhler+18}.
The ASASSN-15lh optical light curve peaked at $\sim 35 \, {\rm d}$ with luminosity $\sim 2 \times 10^{45} \, {\rm erg \, s}^{-1}$ \citep{Dong+16}, and later showed a re-brightening in UV, reaching $\sim 5 \times 10^{44} \, {\rm erg \, s}^{-1}$ at $t \sim 200 \, {\rm d}$ \citep{Brown+16}. Coincidental with the re-brightening, X-rays were first detected from the location of the transient at a luminosity of $\sim 6 \times 10^{41} \, {\rm erg \, s}^{-1}$ (\citealt{Margutti+17b}; there are deeper non-detections at earlier times).

\begin{figure}
\centering
\includegraphics[width=0.45\textwidth]{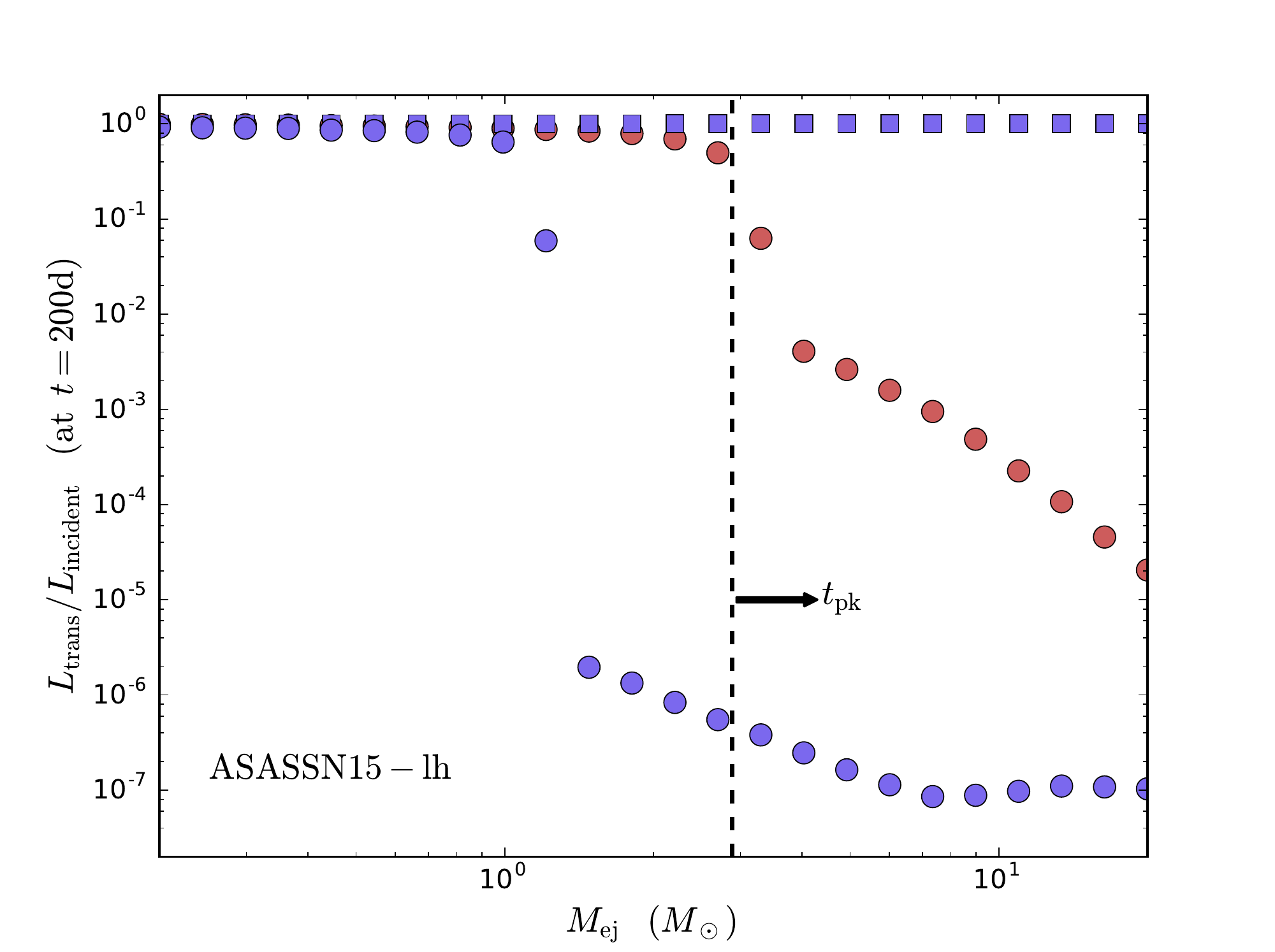}
\caption{
The ratio of transmitted to incident flux in the UV (squares) and X-ray (circles) as a function of assumed ejecta mass for O-rich (blue; SLSN case) and solar composition (red; TDE case). 
The incident luminosity is set to fit the observed UV and X-ray data on ASASSN-15lh at $\sim 200$ days \citep{Brown+16,Margutti+17b}, with a power-law extrapolation of the SED in between these frequency bands.
The incident UV radiation propagates through the ejecta nearly unattenuated for any ejecta mass.  By contrast, the X-ray attenuation is extremely sensitive to the assumed ejecta mass, exhibiting a sharp cut-off above a characteristic ejecta mass, which is $\approx 1 M_\odot$ for the O-rich case and $\approx 3 M_\odot$ for the solar composition case.  The ejecta velocity in both cases is taken to be $v=10^{4} \, {\rm km \, s}^{-1}$, consistent with the observed spectrum of ASASSN-15lh.  A vertical dashed line shows the approximate ejecta mass inferred from the light curve peak under the assumption of a supernova origin for the emission for an assumed opacity $\kappa = 0.2$ cm$^{2}$ g$^{-1}$.} \label{fig:ASASSN15lh}
\end{figure}

One suggested interpretation of the UV (and possibly also X-ray) brightening is ionization break-out of a central-engine, whether the latter is a millisecond magnetar in the SLSN case or an accreting supermassive black hole in the TDE case \citep{Margutti+17b}.  Here we asses whether the detected UV and X-ray luminosities at $t \sim 200 \, {\rm d}$ can be attributed to an ionizing central radiation source behind a layer of expanding ejecta.  We model the engine's incident radiation field by interpolating between the detected luminosities at UV and X-ray frequencies with a power-law SED.\footnote{
We performed a similar calculation but instead modeling the incident SED as two black-bodies of temperatures $1.5 \times 10^4 \, {\rm K}$ and $2 \times 10^6 \, {\rm K}$, respectively, with luminosities necessary to match the X-ray and UV detections \citep{Brown+16,Margutti+17b}. The qualitative results for this spectral model are identical to the power-law SED case, though quantitatively the value of $L_{\rm trans} / L_{\rm incident}$ in Fig.~\ref{fig:ASASSN15lh} at saturation is several orders of magnitude lower.}  This assumption implies that the model will reproduce the observations if the transmitted UV and X-ray luminosities are unattenuated ($L_{\rm trans} \approx L_{\rm incident}$).  Given the unknown mass of the ejecta $M_{\rm ej}$, we explore our results as a function of $M_{\rm ej}$.

Figure~\ref{fig:ASASSN15lh} shows the ratio of transmitted to incident X-ray (circles) and UV (squares) luminosities, separately for O-rich (blue) and solar composition (red) ejecta.  While the UV flux escapes nearly unattenuated for any ejecta mass we explore (unsurprising given the large measured UV luminosity), the X-ray flux show an abrupt step-function transition between being able to ionize 
the ejecta
at low ejecta mass and instead undergoing strong absorption at high $M_{\rm ej}$. The O-rich ejecta exhibits stronger absorption than the Solar composition one due to the larger abundance of bound-free transitions in the X-ray band for this high-metallicity material.
The most striking feature of this result is the nearly bimodal nature of X-ray absorption --- either the incident radiation manages to ionize its way out and escapes nearly unattenuated, or ionization break-out is unsuccessful and the incident radiation is strongly absorbed within the ejecta. A change of only $\sim 50\%$ in ejecta mass can result in six orders of magnitude difference in the escaping X-ray flux (see also Fig.~\ref{fig:x-ray}).

The dashed vertical curve in Fig.~\ref{fig:ASASSN15lh} shows the minimum ejecta mass $M_{\rm ej} \approx 3M_{\odot}$ which is consistent with the observed $35 \, {\rm d}$ optical peak of ASASSN-15lh.  This is made under the assumption that the peak time is determined by the usual photon diffusion timescale $t_{\rm pk} \approx \left( 3 \kappa M_{\rm ej} / 4\pi c v_{\rm ej} \right)^{1/2}$ for an expansion velocity of $v_{\rm ej} = 10^{4} \, {\rm km \, s}^{-1}$ and we take a conservative upper limit on the opacity of $\kappa = 0.2 \, {\rm cm}^2 \, {\rm g}^{-1}$. 

Our results confirm in greater quantitative detail the conclusions of \citet{Margutti+17b}, namely that (1) the observed UV brightening could be the result of an ionization break-out from a central engine, regardless of its precise nature (e.g. a magnetar in the SLSNe case or accreting supermassive black hole in the TDE case); (2) if the observed X-ray emission originates from the same central engine (as opposed to an unrelated source like a nuclear star cluster or AGN), then the ejecta is more consistent with being a TDE than a SLSNe.  For the TDE case the ejecta is expected to be of solar composition and to possess a relatively low mass $\lesssim M_{\odot}$ (most TDEs are expected to be of solar or sub-solar mass stars; e.g.~\citealt{Stone&Metzger16,Kochanek16}).  By contrast, the large mass $\gtrsim 3M_{\odot}$ of oxygen-rich ejecta required in the SLSN scenario would be challenging to photo-ionize through on timescales of the observed X-ray emission.  On the other hand, if the X-rays are unrelated to the transient, as would be the case if they do not fade away in time, then the interpretation would remain ambiguous.  Late-time X-ray observations of ASASSN-15lh to determine if the source has faded would help distinguish these scenarios.

\subsubsection{Pure Fe Ejecta (Binary NS Merger)}

A key question regarding the outcome of binary NS mergers such as GW170817 is whether a central engine, such as a long-lived magnetar or accreting black hole, might contribute to powering or re-energizing the kilonova or afterglow emission \citep{Metzger&Piro14,Metzger&Bower14,Horesh+16,Kisaka+16,Fong+16,Matsumoto+18}.  If present, such an engine could reveal itself through its direct X-ray emission.  Constraints on non-afterglow contributsions to the X-ray emission from GW170817 have been used to argue against the formation of a long-lived NS remnant (\citealt{Pooley+17,Margutti+18}; see \citealt{Margalit&Metzger17} for alternative arguments against a long-lived NS remnant in GW170817).  However, there have thus far been no detailed calculations of the photo-ionization of the merger ejecta and its affect on attenuating a central X-ray source.

\begin{figure}
\centering
\includegraphics[width=0.45\textwidth]{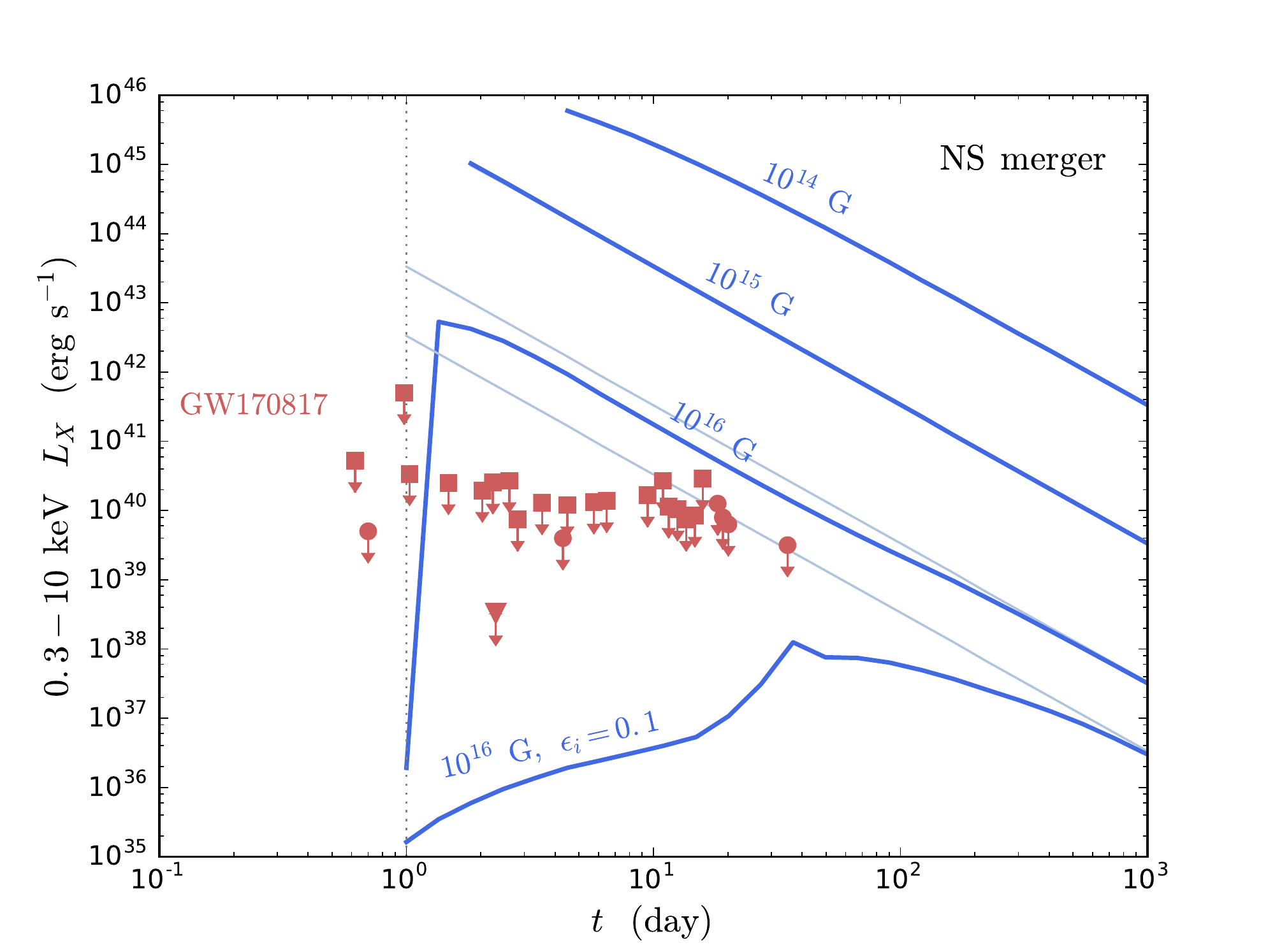}
\includegraphics[width=0.45\textwidth]{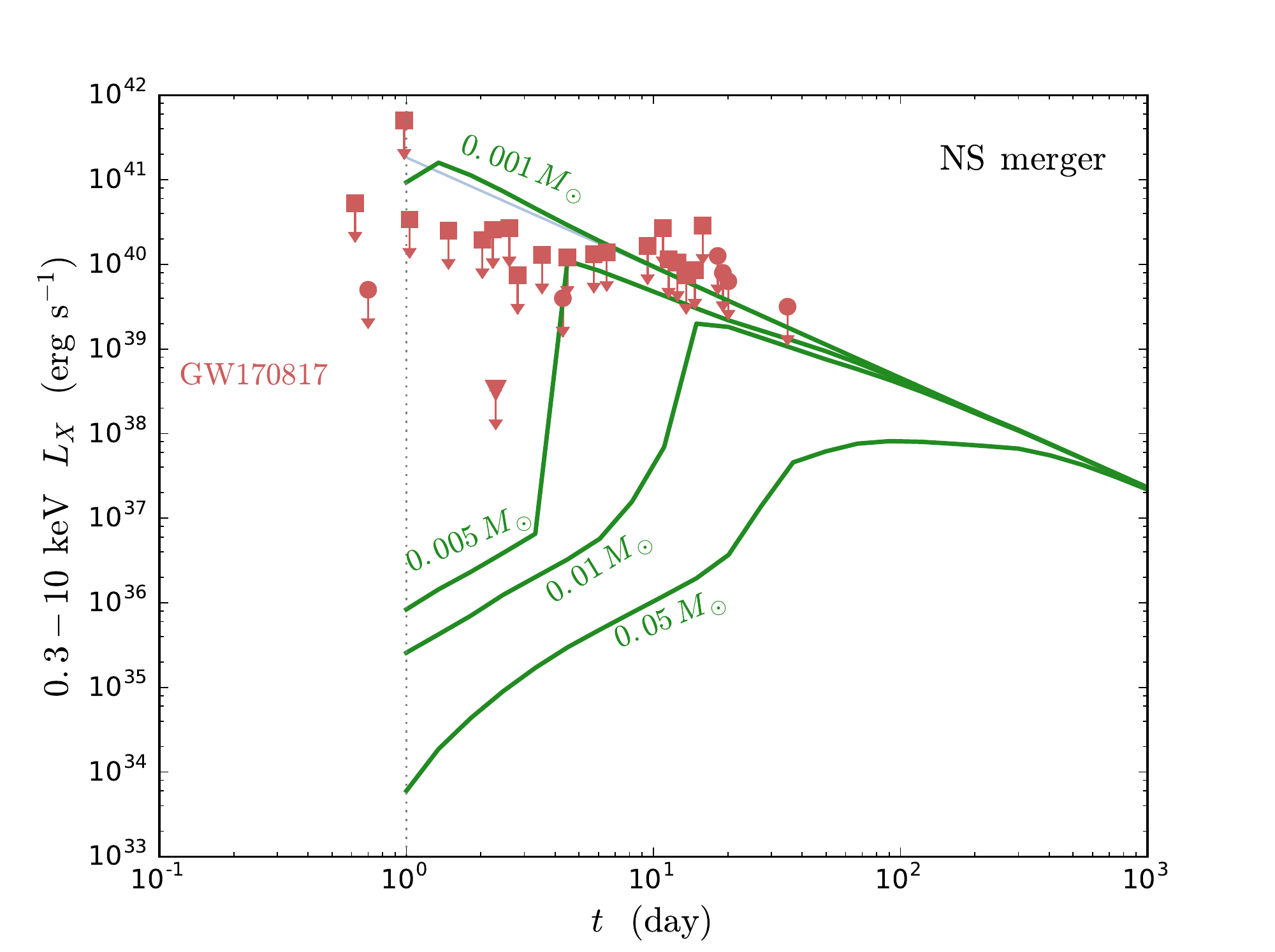}
\caption{X-ray emission from a long-lived magnetar remnant following a binary NS merger.  Similar to Fig.~\ref{fig:x-ray}: unattenuated (light blue) and transmitted (dark blue/green) X-ray light-curves for pure-Fe models of binary NS merger ejecta. The blue curves assume a rotationally-powered magnetar ionizing radiation source of dipole magnetic field strengths 
$B = 10^{14}-10^{16} \, {\rm G}$, while the green curves are for a central engine of ionizing luminosity 
$L_{\rm e} \propto t^{-1.3}$ normalized to the approximate bolometric luminosity of the GW170817 kilonova and different ejecta mass. The red circles, squares and triangle denote GW170817 upper limits from NuSTAR, {\it Swift}, and {\it Chandra X-ray Observatory}, respectively \citep{Evans+17,Margutti+17c}. Clearly, even large $B$ magnetar remnants are ruled out for GW170817. Meanwhile, a more modest luminosity ionizing source is not constrained by the observations.} \label{fig:Fe_Xray}
\end{figure}

Figure \ref{fig:Fe_Xray} shows X-ray light curves resulting from our \cloudy~calculations of expanding ejecta with pure-Fe composition.  Blue curves in the top panel show the transmitted luminosity for a spin-down powered magnetar ionizing source with dipole magnetic fields in the range $B = 10^{14}-10^{16} \, {\rm G}$ and $\epsilon_i = 1$ (except for the bottom curve).  Light-grey curves (in some cases overlapping underneath the blue curves) show for comparison the unattenuated (incident) radiation for each model.  Even high $B$-field spin-down powered engines successfully ionize their way out of the ejecta, such that the transmitted luminosity is nearly equal to the incident one, for large $\epsilon_i$.  A comparison to early-time X-ray upper limits on GW170817 from NuSTAR, {\it Swift}, and {\it Chandra} \citep[red symbols in Fig.~\ref{fig:Fe_Xray};][]{Evans+17,Margutti+17c} rules out magnetar models for this event.

On the other hand for $B = 10^{16} \, {\rm G}$ and a lower radiative efficiency $\epsilon_i=0.1$, X-ray absorption at early epochs is significantly stronger and the presence of a magnetar would be left unconstrained by the X-ray data alone.  This again illustrates the extreme sensitivity of the transmitted flux on model parameters (see also \S\ref{sec:ASASSN15lh}) --- a reduction of the incident flux from $\epsilon_i=1$ to $\epsilon_i=0.1$ results in almost seven orders of magnitude difference in the early-time transmitted luminosity.

The bottom panel of Fig.~\ref{fig:Fe_Xray} shows the transmitted light curve (green curves) through NS merger ejecta, this time 
for a central engine of luminosity $L_{\rm e} = 6 \times 10^{41} \, {\rm erg \, s}^{-1} \left( t/1{\rm d} \right)^{-1.3}$. 
Such an engine has the expected temporal power-law resulting from radioactive decay of $r$-process material to the valley of stability \citep{Metzger+10} and roughly tracks the bolometric luminosity of the kilonova associated with GW170817 \citep[e.g.][]{Arcavi18}.
This second case was chosen to constrain models in which the GW170817 kilonova was powered by a central engine rather than by radioactivity \citep{Matsumoto+18,Li+18}.

Different curves show results for different assumed ejecta masses (labeled for each curve) for fixed ejecta velocity $v_{\rm ej} =0.2c$. For the highest mass cases $M_{\rm ej} \gtrsim 10^{-2} M_\odot$, the ejecta provides a sufficient column density to absorb the X-rays at early times, consistent with GW170817 non-detections.  However, lower mass ejecta, such as $M_{\rm ej} \sim 10^{-3}M_{\odot}$ advocated by \cite{Li+18}, are ruled out.  We stress that for $M_{\rm ej} \gtrsim 10^{-2} M_\odot$, the $r$-process radioactive heating rate becomes comparable to the assumed engine luminosity, essentially bypassing the need for invoking such an engine.

\subsection{Radio Transparency and Ejecta DM} \label{sec:DM_freefree}

Having calculated both the ionization and temperature structure of the ejecta for an ensemble of properties motivated by SLSNe, we now determine the local DM contribution to the ejecta and its free-free absorption.  The latter controls both when a putative FRB and associated nebular emission produced within such SLSNe ejecta become visible.

The DM is naturally expressed in terms of the density-averaged ejecta ionization fraction
\begin{equation}
{\rm DM} = \int n_{\rm e} \, dr
= \frac{9 M_{\rm ej}}{20 \pi \mu_{\rm e} m_{\rm p} v_{\rm ej}^2 t^2} \left\langle f_{\rm ion} \right\rangle_\rho(t) ,
\end{equation}
where $\mu_{\rm e} = \rho/(m_p n_{\rm e}) \simeq 2$ is the mean molecular weight per electron.

\begin{figure}
\centering
\includegraphics[width=0.45\textwidth]{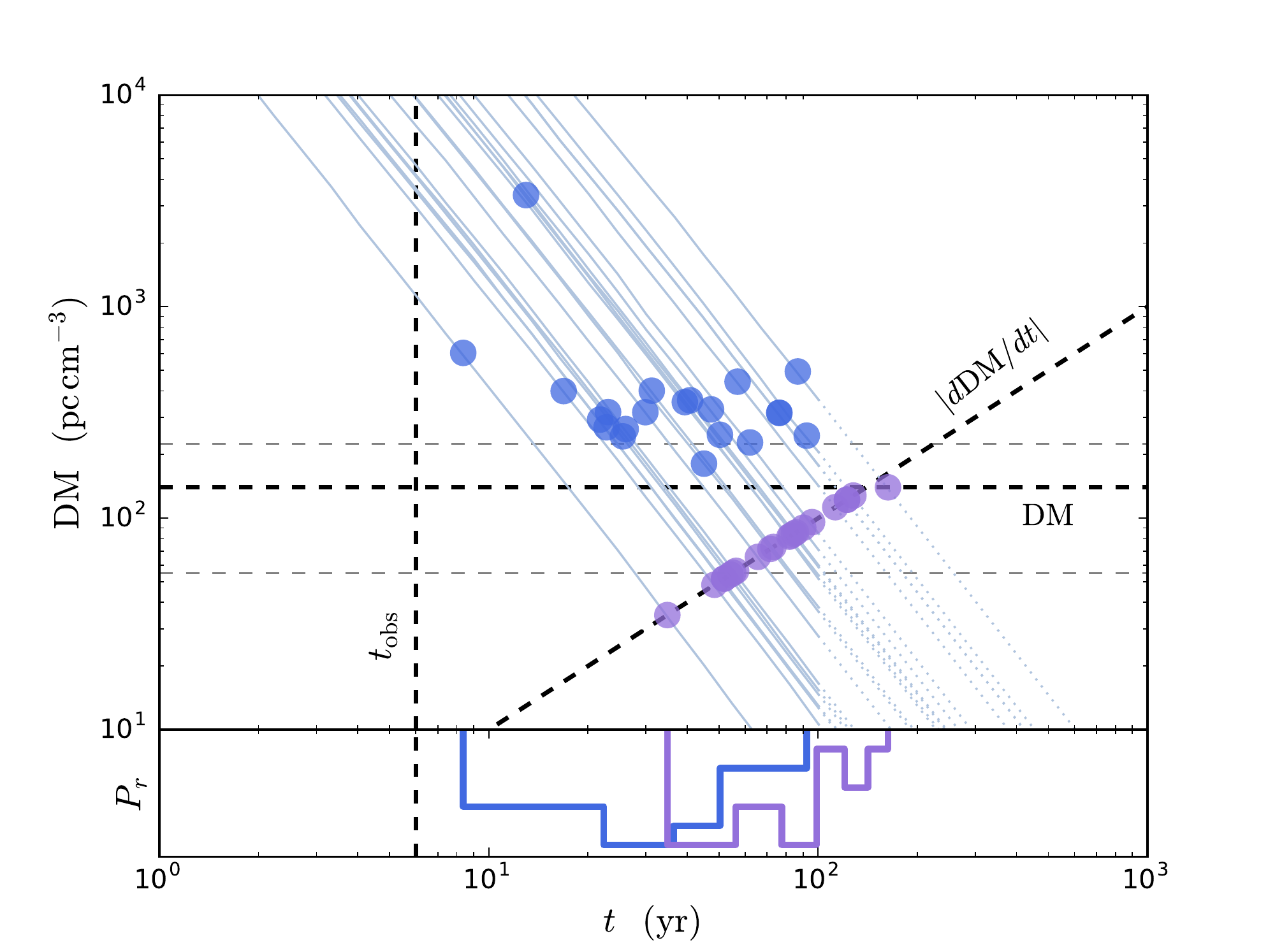}
\caption{Time evolution of the ejecta dispersion measure ${\rm DM}(t)$ for the SLSNe in our sample (solid grey curves; extrapolated by dotted extensions).  Blue points show the time at which the ejecta first becomes transparent to free-free absorption at $1 \, {\rm GHz}$.  Observational constraints for FRB 121102 are shown as dashed black curves, whose intersection with the ${\rm DM}(t)$ tracks indicate the minimum allowed age of FRB 121102 assuming it originates from each SLSN (purple points). The bottom panel shows the distribution of free-free transparency timescales and minimal age (blue and purple respectively). If FRB 121102 originates from a population of young magnetars with properties similar to those inferred for observed SLSNe, then its age is $\gtrsim 30-100 \, {\rm yr}$.} \label{fig:DM_t}
\end{figure}

Figure \ref{fig:DM_t} shows the time evolution of the ${\rm DM}$ for the population of SLSNe in our sample for $\epsilon_{i} = 1$ (solid grey curves). Since we found that the ionization fraction is nearly constant in time  (\S\ref{sec:ionization_state}), the DM evolution closely follows a simple $\propto t^{-2}$ power-law decay due to the decreasing ejecta column, as shown by the dotted-grey extrapolations to late times after conclusion of the \cloudy~calculations. Various tracks are therefore differentiated almost entirely based on their different normalizations imprinted by the ejecta mass and ionizing spin-down power.

The free-free optical depth can be approximately expressed as 
\begin{align}
\tau_{\rm ff} &= 
\int \kappa_0 g_{\rm ff}(T_{\rm e}, \nu) \nu^{-2} Z^2 n_e n_i T_{\rm e}^{-3/2} dr
\\ \nonumber
&\approx \kappa_0 g_0 \nu^{-2.118} Z^{1.882} \int T_{\rm e}^{-1.323} f_{\rm ion}^2 n^2 dr ,
\end{align}
where the gaunt factor is $g_{\rm ff} \approx g_0 ( Z \nu )^{-0.118} T^{0.177}$ \citep{Draine11}, and $g_0 = 13.907$, $\kappa_0 = 0.01772$ in appropriate cgs units.
Neglecting temperature changes and for a $\sim$constant $\left\langle f_{\rm ion} \right\rangle_\rho$, this implies a temporal scaling of $\tau_{\rm ff} \propto t^{-5}$. 
From the numerical \cloudy~calculations, 
we find $d \ln \tau_{\rm ff} / d \ln t \sim -4.2$ to $-4.6$ up to the $1 \, {\rm GHz}$ transparency timescale $t_{\rm ff} \sim 10-100 \, {\rm yr}$ for the SLSNe in our sample.

\subsubsection{Comparison to the Reverse Shock}
\label{sec:reverse}
The SN/merger ejecta will become collisionally-ionized after being heated by the reverse shock, which propagates back into the ejecta as the ejecta interacts with the circumstellar medium \citep{Piro16}.  This process occurs on the Sedov-Taylor timescale,
\begin{equation} \label{eq:t_ST}
t_{\rm ST} \approx 450 \, {\rm yr} \left(\frac{E}{10^{52} \, {\rm ergs}}\right)^{-1/2} \left(\frac{M_{\rm ej}}{10 M_\odot}\right)^{5/6}\left(\frac{n_{\rm csm}}{1\,{\rm cm}^{-3}}\right)^{-1/3} ,
\end{equation}
where $n_{\rm csm}$ is the number density of the external medium.  Here we compare this source of external shock-ionization to that due to photo-ionization from the central engine.

\begin{figure}
\centering
\includegraphics[width=0.45\textwidth]{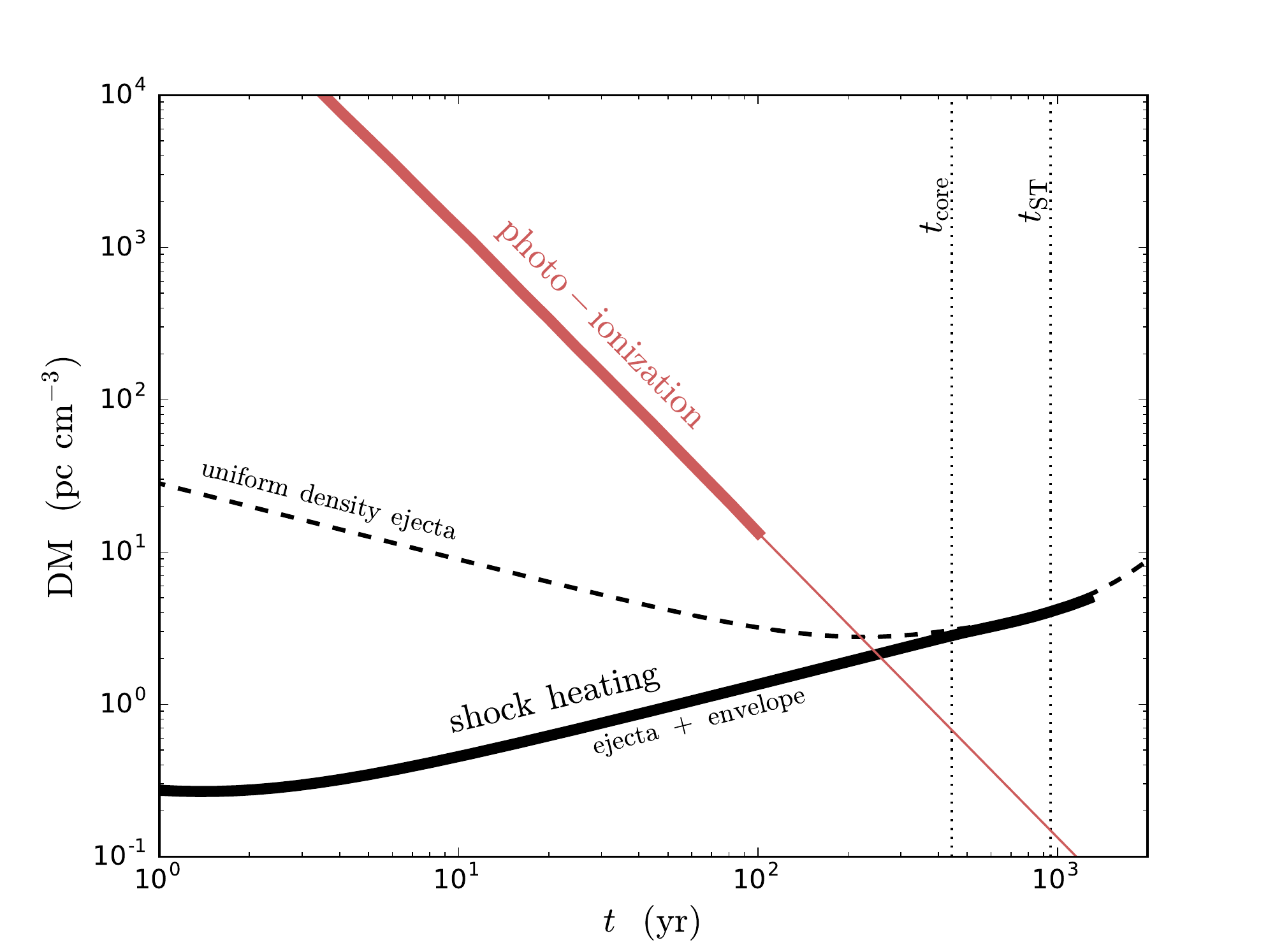}
\caption{Comparison of the DM contribution from the reverse shock driven into the ejecta by interaction with the circumstellar medium (black curves) with that of the DM contribution from photo-ionization due to a central engine (red curve), as presented previously in Figure \ref{fig:O_IonizationState}.  The dashed black curve shows the DM for the idealized case of homogeneous ejecta \citep{Piro16}, while the solid black line shows results for the more realistic case of a steeply-declining outer envelope (equation~\ref{eq:rho_SN}), 
which exhibits a much lower DM at early times when the reverse shock is propagating within the low-density envelope.  As expected, the two solutions converge once the shock enters the ejecta core (dotted-vertical curve marked by $t_{\rm core}$) and the dynamics transition to that of the Sedov-Taylor phase (second dotted-vertical line, $t_{\rm ST}$).  In either case, the DM contributed
by the forward/reverse shocks is significantly lower than that due to photo-ionization by a central engine on timescales $\lesssim 10^2$ yr.}
\label{fig:reverseshock_DM}
\end{figure}

\cite{Piro16} assume a constant density ejecta and find, using the approximate hydrodynamic solutions of \cite{McKee&Truelove95}, a large ejecta DM contribution at early times $t\ll t_{\rm ST}$ from the reverse shock.  However, more realistic models for the ejecta structure find steeply-declining outer envelopes outside of the relatively flat, constant-density core (e.g.~\citealt{Chevalier&Soker89,Suzuki&Maeda17}; see equation~\ref{eq:rho_SN}), which exhibit dramatically different behavior at early times, $t \lesssim t_{\rm ST}$.  Using the approximate solutions of \cite{Truelove&McKee99} we re-evaluate this contribution to the local DM and ionization-state (see Appendix \ref{sec:Appendix_reverseshock}).

Figure \ref{fig:reverseshock_DM} shows our results for the time-dependent DM contribution from the reverse shock for a characteristic\footnote{We obtain similar results for other values $n \gtrsim 5$, although the limit of $n \to \infty$ reverts back to the case of homogeneous ejecta \citep{Piro16}.} ejecta profile (equation~\ref{eq:rho_SN}; $n=6$ in the notation of \citealt{Truelove&McKee99}), and compared to the idealized constant-density ejecta distribution ($n=0$) as well as the fiducial O-rich photo-ionization model presented previously in Figure \ref{fig:O_IonizationState}.  In both models photo-ionization dominates the DM at $t \lesssim 200 \, {\rm yr}$, while the reverse shock dominates at later times.  Although the case shown assumes typical SLSNe parameters ($E = 10^{52} \, {\rm ergs}$; $M = 10 M_\odot$; and $n_{\rm csm} = 1 \, {\rm cm}^{-3}$), the time axis can be scaled trivially with $t_{\rm ST}$ (equation~\ref{eq:t_ST}).  Our DM estimate does not account for details of the post-shock density distribution due to compression or radiative cooling, which however will only act to reduce the ionized column by a factor of $\lesssim 3$ (Appendix \ref{sec:Appendix_reverseshock}). Furthermore, for $n=6$ the majority contribution to the ionized mass and DM comes from the shocked circumstellar medium instead of the shocked ejecta.

The main physical effect leading to the smaller DM 
when the ejecta has a steep outer envelope compared to the constant density ejecta
\citep{Piro16} is the larger blast-wave and reverse shock radii at a given time (and thus smaller column for a fixed ejecta mass) in the envelope case,
since at early times the forward and reverse shocks are located at velocity coordinates greater than $v_{\rm ej}$ (at which the core-envelope transition occurs).

\subsubsection{Application to FRB~121102}

Observational constraints on the local contributions of the DM and its time derivative for FRB 121102, as well as the system age, are given by (e.g.~\citealt{Spitler+16,Piro16,Chatterjee+17})
\begin{align}
&{\rm DM}_{\rm local} \lesssim 140 \pm 85 \, {\rm pc \, cm}^{-3} ,
\\
&\left\vert d{\rm DM}/dt \right\vert \lesssim 2 \, {\rm pc \, cm}^{-3} \, {\rm yr}^{-1} ,
\\
&t_{\rm age} \gtrsim 6 \, {\rm yr} ,
\end{align}
and are depicted as dashed black curves in Fig.~\ref{fig:DM_t}.
The ejecta first becomes optically thin to free-free absorption (blue points) at $t_{\rm ff} \sim 10-100 \, {\rm yr}$ after explosion (see histogram on the bottom panel), while the minimal age of FRB~121102 (purple points), assuming it originates from a magnetar with properties characteristic of the SLSN population, is $\sim 30-100 \, {\rm yr}$.
Our detailed numerical result is thus consistent with previous analytic estimates of the repeater's minimal age \citep{Metzger+17}.

Figure \ref{fig:DM_t} illustrates that the constraints on FRB 121102 are consistent with all of the SLSNe in our sample at sufficiently late times.  However, this alone says nothing about the {\it probability} we are observing the repeater at such a late time.  To explore this issue, we assume that the probability distribution for detecting an FRB at time $t$ is given by $\Prob \left( t \vert t_{\rm ff} \right) \propto t^{-\alpha}$, provided that $t > t_{\rm ff}$, the free-free transparency timescale, and $t < t_{\rm a}$, the engine activity timescale.  While $t_{\rm ff}$ is determined by our \cloudy~calculations, the parameter values $\alpha$ and $t_{\rm a}$ which control the rate of FRB activity are uncertain (though $t_{\rm a}$ is expected to be less than a few hundred years if the flares result from the diffusion of magnetic flux from the magnetar core; $\S\ref{sec:magnetic}$).  For $\alpha$ we consider two cases: (i) a `flat' evolution ($\alpha=0$), motivated by magnetic dissipation powered FRB models (equation~\ref{eq:Lmag}); and (ii) a rapidly decaying evolution $\alpha=2$ model, e.g.~motivated by spin-down powered FRB scenarios (equation~\ref{eq:Lrot}).

Since the DM undergoes a simple time evolution $\propto t^{-2}$, it is easy to invert the problem to calculate the probability of observing an FRB from a specific SLSN with some DM. Summing over the distribution of $t_{\rm ff}$ and the DM at this time for the population of SLSNe then yields the marginalized probability distribution of observing an FRB with some given dispersion measure, $\Prob \left( {\rm DM} \right)$ (see Appendix~\ref{sec:Appendix_P_of_DM} for further details).
The same procedure can be repeated for the DM derivative, allowing comparison with constraints on FRB 121102.

\begin{figure*}
\centering
\includegraphics[width=0.45\textwidth]{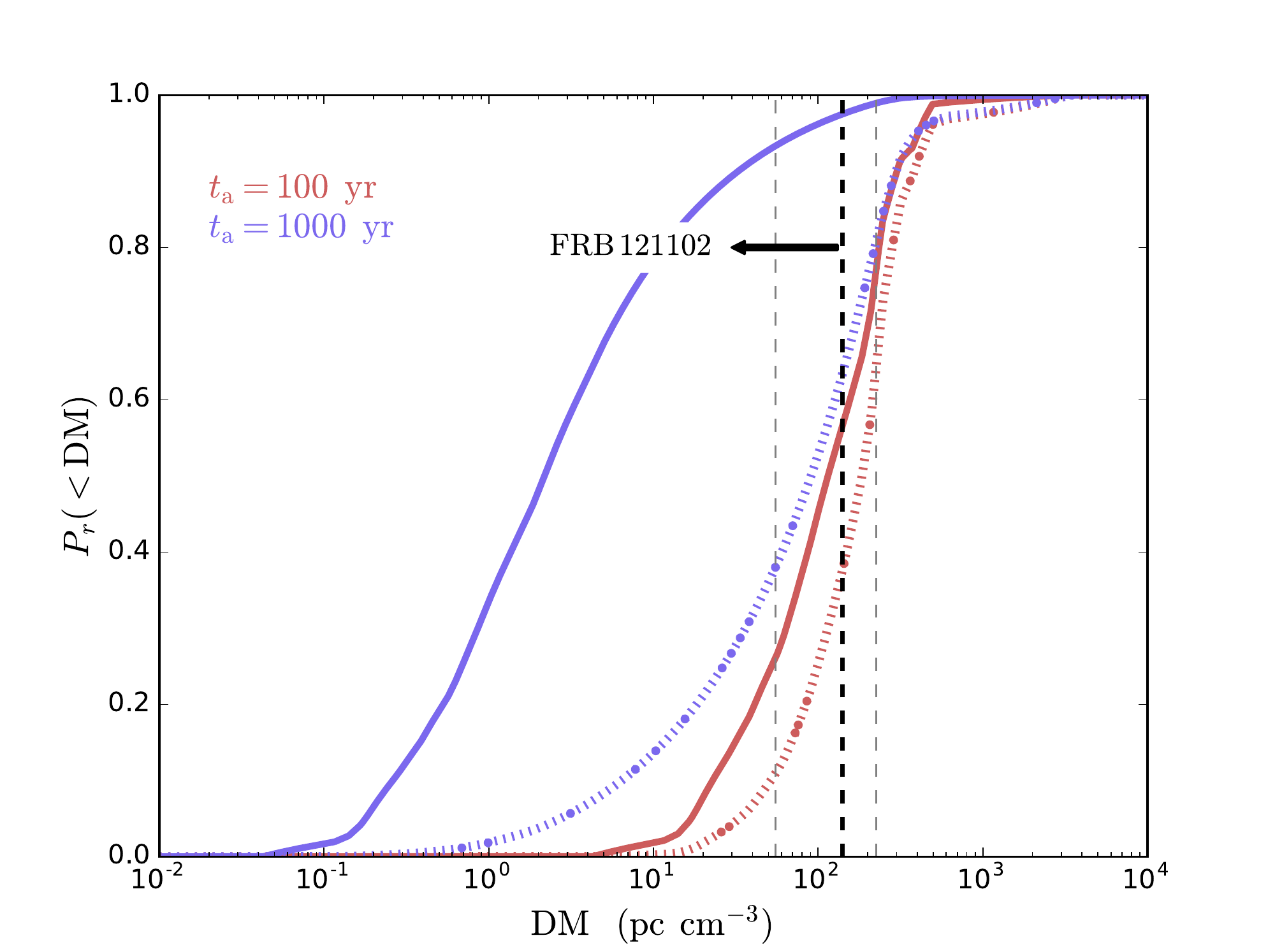}
\includegraphics[width=0.45\textwidth]{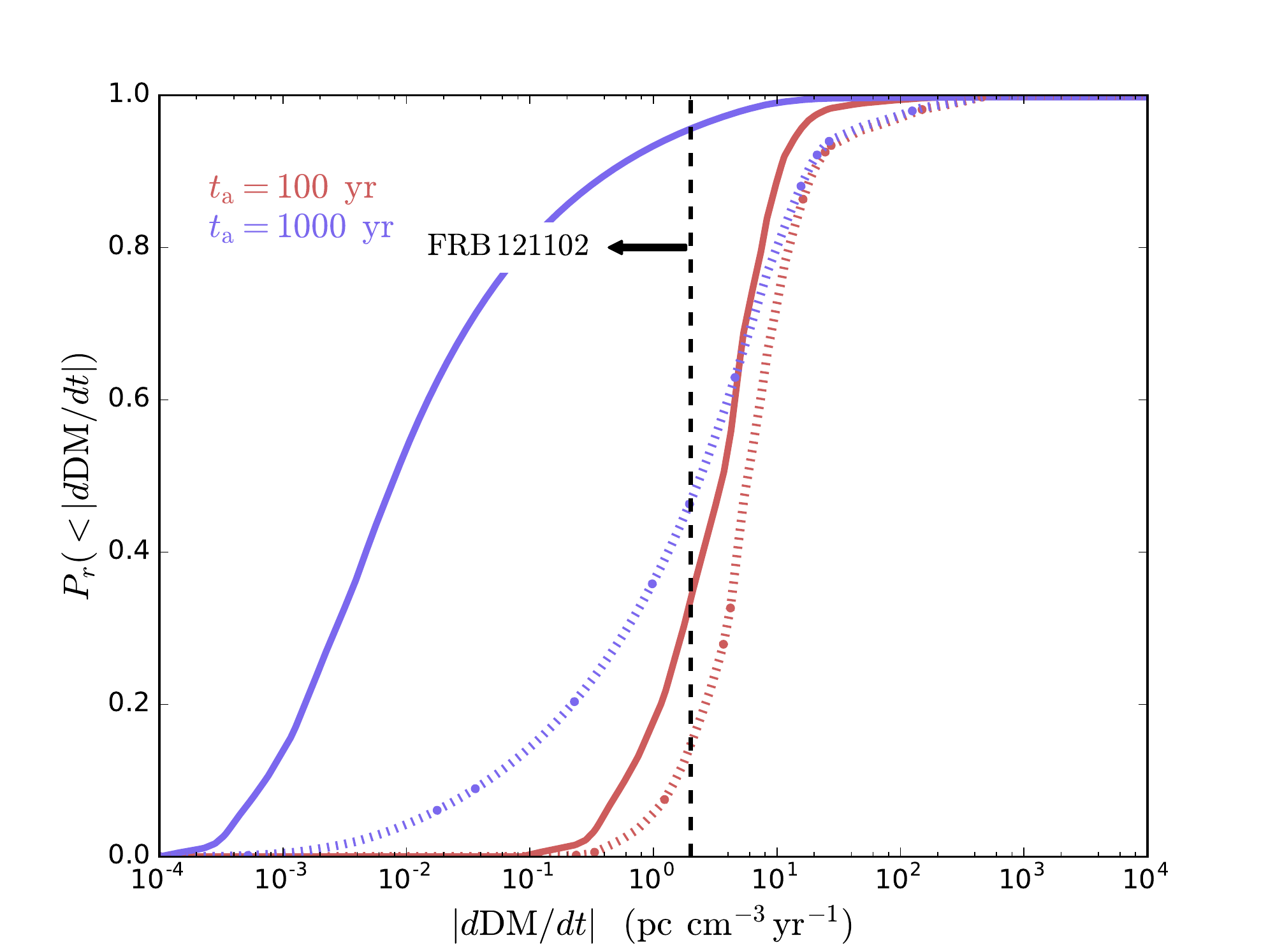}
\caption{Probability distributions of observing an FRB with a given local dispersion measure and DM derivative, assuming FRBs are produced by a population of magnetars (and their enveloping SN ejecta) similar to the observed SLSN population.
The probability distribution functions are calculated based on the DM and free-free transparency times found using our \cloudy~models, and assuming an FRB activity lifetime $t_{\rm a}$ (different colored curves) and an FRB activity / detection metric which evolves with time as $t^{-\alpha}$ (see \S\ref{sec:DM_freefree} and Appendix~\ref{sec:Appendix_P_of_DM} for further details). Solid curves show results for $\alpha=0$ while dotted curves are for $\alpha=2$.
The observational constraints (upper-limits) for FRB~121102 are shown as dashed vertical curves. The probability that the local DM and $d{\rm DM}/dt$ of FRB~121102 be consistent with predictions for the population of SLSNe is clearly non-negligible for a wide range of model parameters. The hypothesis that FRB~121102 arises from a young magnetar with parameters ($P_0,B,M_{\rm ej}$) drawn from the observed SLSN population is therefore consistent with the observed dispersion measure and its time derivative.}
\label{fig:probability_distributions}
\end{figure*}

Figure~\ref{fig:probability_distributions} shows the resulting probability distributions for $\alpha=0$ (solid curves) and $\alpha=2$ (dotted curves), and different assumptions regarding the engine active lifetime, $t_{a}$ (different colors).  Shown for comparison are the constraints on DM and its derivative for FRB 121102. 
In practice, as in Fig.~\ref{fig:DM_t}, the upper limit on $d{\rm DM}/dt$ provides the tightest constraints.   
In the fiducial, magnetically powered model ($\alpha=0$), even a relatively short engine lifetime of $t_{\rm a}=100 \, {\rm yr}$ results in a non-negligible probability $\Prob (d{\rm DM}/dt < 2 \, {\rm pc \, cm}^{-3} \, {\rm yr}^{-1} ) = 0.34$ of observing FRB~121102 at times consistent with the constraints.  A longer assumed engine activity timescale obviously results in a higher probability of detecting the source at sufficiently late times.  The rapidly decaying model ($\alpha=2$) predicts somewhat lower, yet still significant, probabilities of randomly detecting an FRB with ${\rm DM}$ and $d{\rm DM}/dt$ consistent with the repeater. At the low end, for $t_{\rm a}=100 \, {\rm yr}$, we obtain $\Prob (d{\rm DM}/dt < 2 \, {\rm pc \, cm}^{-3} \, {\rm yr}^{-1} ) = 0.15$, while even slightly longer timescales approach the asymptotic $t_{\rm a} \to \infty$ limit of $\Prob (d{\rm DM}/dt < 2 \, {\rm pc \, cm}^{-3} \, {\rm yr}^{-1} ) = 0.47$.

In summary, we conclude that, under a wide range of assumptions, the observed DM and $d{\rm DM}/dt$ of the repeater are completely ``characteristic'' of those expected if FRB~121102 originates from an engine embedded within a young SLSN.

\section{Nebula Rotation Measure and Synchrotron Emission}
\label{sec:radio}

In the previous section we presented calculations of the time-dependent ionization state of the SN ejecta, in order to predict its DM and the X-ray light curves from the central engine.  This section extends this connection to the RM and the radio emission from synchrotron nebulae with relation to the quiescent radio source associated with FRB~121102 \citep{Chatterjee+17,Marcote+17}.  Given the ejecta's free-free optical depth calculated in $\S\ref{sec:DM_freefree}$, we also estimate the late-time radio emission from SLSNe and long GRBs, and from magnetar-powered FRB sources more generally.

\subsection{Rotation Measure}
\label{sec:RM}

The large RM $\approx 10^{5}$ rad m$^{-2}$ of FRB~121102 \citep{Michilli+18}, and its observed $\sim$ 10\% decrease over a baseline of 7 months, if related to the dillution of an expanding nebula, strongly constrain the age and origin of the bursting source. 

\subsubsection{An Electron-Ion Nebula}

Since the RM contributions from positrons and electrons cancel one another, the observed large RM value requires an electron-ion plasma rather than the pair-dominated ultra-relativistic wind from a rotationally-powered pulsar wind \citep{Michilli+18}.  Though young pulsars produced primarily electron-positron winds,  
a large ion loading is not necessarily surprising in the context of a bursting magnetar.  Observations of the synchrotron radio afterglows of giant flares from Galactic magnetars indeed find the bulk of the matter ejected from these events to be expanding at mildly- or trans-relativistic speeds (e.g.~\citealt{Granot+06}).  This substantial baryon loading is presumably from the neutron star surface after being heated during the fireball phase (following which most electron/positron pairs annihilate; e.g.~\citealt{Beloborodov17}).  

A key, but theoretically uncertain, property of the magnetically-powered outflow is the average ratio, $\xi$, of the number of ejected baryons to the released magnetic energy \citep{Beloborodov17}
\begin{equation}
10^{2}\,{\rm erg^{-1}} \lesssim \xi \equiv \frac{N}{E} \lesssim 10^{4}\,{\rm erg^{-1}}.
\label{eq:xi}
\end{equation}
Here the lower limit on $\xi$ follows from an estimate of the minimum number of radio-emitting electrons responsible for the afterglow of the giant flare of SGR 1806-20 \citep{Granot+06}, while the upper limit corresponds to the escape speed of a neutron star,
\begin{equation}
\xi_{\rm max} \approx \frac{R_{\rm ns}}{GM_{\rm ns}m_{\rm p}} \approx 4\times 10^{3}\,{\rm erg^{-1}}.
\label{eq:ximax}
\end{equation}

If the kinetic energy of the ejecta flare thermalizes at the termination shock, transfering a fraction $\epsilon_{e}$ of its energy to the electrons, then the latter enter the nebula with a mean thermal Lorentz factor
\be
\bar{\gamma}_e \approx \frac{\epsilon_e}{\xi} \approx 150\left(\frac{\epsilon_{e}}{0.5}\right)\left(\frac{\xi}{\xi_{\rm max}}\right)^{-1}.
\label{eq:gammabar}
\ee
As discussed below, the characteristic value $\bar{\gamma}_e \sim 10^{2}$ implied for $\xi \lesssim \xi_{\rm max}$ is consistent with that required to power the quiescent sychrotron source from FRB~121102.  The limited frequency range $\nu \sim 1-20$ GHz over which the quiescent source is observed, and the potential impact of cooling on the spectrum, makes it challenging to determine whether the radiating electron population is a non-thermal power-law (as assumed in previous works), or whether it might be also consistent with a relativistic Maxwellian (or superposition of Maxwellians) with $kT \approx \bar{\gamma}_{e} m_e c^{2}$ (equation~\ref{eq:gammabar}).

\subsubsection{Radio-Emitting Electrons, Injected Recently} 

We first show that FRB~121102's high RM cannot originate from the same relativistic electrons responsible for powering the quiescent radio emission.  Synchrotron emission from electrons with Lorentz factor $\gamma_e = 100\gamma_{100}$ embedded in a magnetic field $B$ (in Gauss) peaks at a frequency $\nu \approx 5.6B\gamma_{100}^{2}\,{\rm GHz}$.  The observed spectral luminosity $L_{\nu} \approx 10^{29}\nu_{10}^{-0.2} \, {\rm erg \, s}^{-1} \, {\rm Hz}^{-1}$ of the source at $\nu < 10$ GHz is related to the number of radiating electrons $N_{\gamma_e} \equiv dN_{\gamma_e}/d {\rm ln\gamma_e}$ with $\gamma_e(\nu)$ according to \citep{Beloborodov17}
\be
L_{\nu} \approx 3 \frac{e^{3}B}{m_e c^{2}}N_{\gamma_e} \Rightarrow N_{\gamma_e}B \approx 2\times 10^{50}{\rm G}.
\ee For a homogeneous spherical nebula of radius $R_{\rm n} = 10^{17} R_{17} \, {\rm cm}$, and magnetization parameter $\sigma$ (ratio of magnetic to particle energy), one finds individually that \citep{Beloborodov17}
\be
B \approx 0.06\sigma^{2/7}R_{17}^{-6/7}\,{\rm G}; \\
N_{\gamma_e} \approx 3\times 10^{51}\sigma^{-2/7}R_{17}^{6/7},
\ee
and thus the Lorentz factor of the emitting particles is
\be
\gamma_{e} \approx 540 \nu_{10}^{1/2}\sigma^{-1/7}R_{17}^{3/7},
\label{eq:gamma}
\ee
while their average number density in the nebula is
\be
n_{e} \approx \frac{3N_{\gamma_e}}{4\pi R_{\rm n}^{3}} \approx 0.7\,{\rm cm^{-3}}\,\,\sigma^{-2/7}R_{17}^{-15/7}.
\ee
The maximum RM through the nebula, from the same electrons which power the observed synchrotron radiation, is then given by
\begin{eqnarray}
{\rm RM_{\gamma_e}} &=& \frac{e^{3}}{2\pi m_e^{2}c^{4}}\int \frac{n_{\rm e}}{\gamma_e^{2}{\rm ln \gamma_e}}B_{\parallel} ds \lesssim \frac{e^{3}}{2\pi m_e^{2}c^{4}}\frac{n_{\rm e}B R_{\rm n}}{\gamma_e^{2}{\rm ln \gamma_e}},  \nonumber \\
&\approx& 6\times 10^{-4}{\rm rad\,m^{-2}}\sigma^{2/7}R_{17}^{-20/7}
\end{eqnarray}
where the $1/(\gamma_e^{2}{\rm ln \gamma_e})$ factor accounts for suppression of the RM contributed by relativistically-hot electrons (e.g.~\citealt{Quataert&Gruzinov00}) and in the second line we have neglected the parameter dependence of the logarithmic terms.  Clearly, the value of ${\rm RM}_{\gamma_e}$ from the radio-emitting electrons is many orders of magnitude too low to explain the observed RM $\sim 10^{5}$ rad m$^{-2}$.

\subsubsection{Cooled Electrons, Injected in the Distant Past}

Recently-injected electrons responsible for the quiescent radio source of FRB~121102 cannot produce its large RM, in part because of the $\propto 1/\gamma_{e}^{2}$ suppression for relativistic temperatures.  However, prospects are better if a greater number of electrons were ejected when the source was younger, especially  since they may by now be sub-relativistic ($\gamma_e \lesssim 2$) due to synchrotron and adiabatic cooling.  
Such cooling is reasonable if the present source age is $\gtrsim 3-10$ times greater than the timescale $t_{\rm mag}$ around when magnetic activity peaked (equation~\ref{eq:Lmag}) and presumably when most of the baryons were deposited in the nebula.  Adiabatic expansion alone will reduce the energy of relativistic electrons by a factor $\sim t/t_{\rm mag}$ from their injected ultra-relativistic values, while the synchrotron loss timescale $\propto 1/B_{\rm n}^{2} \propto t^{2}\dot{E}_{\rm mag}$ will also be considerably shorter at early times when the magnetic field inside the nebula is stronger (see equation~\ref{eq:Bn} below).   

To explore this possibility with a rough estimate, consider that the magnetar has up until now released a magnetic energy $E \sim E_{\rm B} = 10^{51} E_{51} \, {\rm erg}$, comparable to its total magnetic reservoir $E_{\rm B}$ (equation~$\ref{eq:EB}$).  The total number of electrons in the nebula is therefore (equation~\ref{eq:xi})
\be
N_{e}  = \xi E_{\rm B} \approx 4\times 10^{54}(\xi/\xi_{\rm max})E_{51}
\ee
where a value $\xi \sim \xi_{\rm max}$ (equation~\ref{eq:ximax}) is again motivated by matching the thermal Lorentz factor of the injected electrons (equation~\ref{eq:gammabar}) to those required to explain the frequency of the quiescent radio emission of FRB~121102 (equation~\ref{eq:gamma}).  The average density of electrons in the nebula is then $n_{e} \approx 3N_{\rm e}/(4\pi R_{\rm n}^{3})$. 

If the magnetic energy of the nebula, $B_{\rm n}^{2}R_{\rm n}^{3}/6$, is a fraction $\epsilon_{B}$ of the energy $\sim L_{\rm mag} t$ injected in relativistic particles over an expansion time $\sim t$, then the magnetic field strength in the nebula is given by
\be
B_{\rm n} \approx \left(\frac{6\epsilon_{B}E_{\rm B} (\alpha-1)}{R_{\rm n}^{3}}\right)^{1/2} \left(\frac{t}{t_{\rm mag}}\right)^{(1-\alpha)/2},
\label{eq:Bn}
\ee
 where we have used equation (\ref{eq:Lmag}) for $L_{\rm mag}(t)$. 

Combining results, the maximum contribution to the RM (assuming all the electrons are mildly relativistic) is given by
\begin{eqnarray}
&&{\rm RM} = \frac{e^{3}}{2\pi m_e^{2}c^{4}}\int n_e B_{\parallel} ds \approx  \frac{3e^{3}}{8\pi^{2} m_e^{2}c^{4}} \frac{N_{\rm e}B_{\rm n}}{R_{\rm n}^{2}}\left(\frac{\lambda}{R_{\rm n}}\right)^{1/2}\nonumber \\
&\approx& 6\times 10^{7} [\epsilon_{B}(\alpha-1)]^{1/2}E_{51}^{3/2}\left(\frac{\xi}{\xi_{\rm max}}\right)\times \nonumber \\
&&R_{17}^{-7/2}\left(\frac{t}{t_{\rm mag}}\right)^{\frac{(1-\alpha)}{2}}\left(\frac{\lambda}{R_{\rm n}}\right)^{1/2}\,{\rm rad\,m^{-2}} \nonumber \\
\end{eqnarray}
where $\lambda$ quantifies the correlation lengthscale of the magnetic field in the nebula.  Thus, we see it is possible to obtain RM values $\sim 10^{5}$ rad m$^{-2}$ comparable to those measured for FRB121102 for optimistic parameters, e.g. $\epsilon_{B} = 0.1$, $E_{51} \approx 1$, $t = t_{\rm age} \sim 10 t_{\rm mag}$, $\lambda \sim R_{\rm n}$.

An important prediction of this model is the expected secular decrease in the RM.  Assuming that $R_{\rm n} \propto t$, the time derivative of the RM is given by
\be
\frac{d{\rm RM}}{dt} = -\frac{(6+\alpha)}{2}\frac{RM}{t}
\ee
If the observed $\Delta {\rm RM}/{\rm RM} = -0.1$ change in FRB~121102's RM over the baseline of $\Delta t = 0.6$ yr \citep{Michilli+18} is entirely due to this secular decline, then this requires a source age of
\be t = t_{\rm age} \approx \frac{\alpha+6}{2}\left(\frac{RM}{\Delta RM}\right)(\Delta t) \approx 5(\alpha+6)\,\,{\rm yr}
\label{eq:RMage} \ee  
Instead treating the observed change as an {\it upper limit} on the change in the RM due to secular expansion (e.g.~if the observed variability is dominated by some stochastic process, e.g. internal turbulence, at fixed nebular size), then equation (\ref{eq:RMage}) becomes a lower limit on the source age.
This RM-inferred age estimate is compatible with the $\sim 30-100 \,{\rm yr}$ age independently estimated based on the DM-derivative.

\subsection{Radio Synchrotron Emission}

Though X-rays from the engine appear challenging to detect (Fig.~\ref{fig:x-ray}), prospects may be better at radio frequencies once the ejecta becomes transparent to free-free absorption \citep{Kashiyama+16}.  Indeed, such a nebula was proposed as the origin of the quiescent radio source associated with FRB~121102 \citep{Metzger+17,Kashiyama&Murase17}.  If FRB~121102 is indeed associated with a young magnetar, its birth heralded by a SLSN or long GRB, then one may invert the problem to ask what radio emission we might expect to detect at late times from other SLSNe or long GRB remnants \citep{Metzger+17}.  In the following, we adopt a phenomenological approach to estimating the late-time quiescent radio flux, using a minimal set of assumptions and scaling whenever possible to the observed properties of the repeater's quiescent source.

We consider two assumptions about the energy spectrum of electrons injected into the nebula which radiaion synchrotron emission.  First, we assume that the electrons are injected with a relativistic Maxwellian of constant temperature (or, equivalently, mean Lorentz factor) given by equation (\ref{eq:gammabar}). We also consider a power-law population of electrons accelerated in the magnetar nebula to Lorentz factors $\gamma$, $\partial N / \partial \gamma \propto \gamma^{-p}$, which emit at frequencies 
$\nu = \gamma^2 e B / 2 \pi m_e c$. 
We additionally assume that the nebula is observed at early times in the fast-cooling regime ($\nu_{\rm c} < \nu$) such that the radio luminosity is
\begin{equation} \label{eq:Lnu_fastcool}
\nu L_{\rm e,\nu} \propto \dot{E}_{\rm n} \gamma^2 \left( \partial N / \partial \gamma \right) ,
\end{equation}
where $\dot{E}_{\rm n}(t)$ energy injection rate into the nebula from the engine, and the synchrotron spectrum is $L_{\rm e,\nu} \propto \nu^{-p/2}$. 
Finally, we assume that the nebula magnetic field is in equipartition, such that $B \propto \left( E_{\rm n} / R_{\rm n}^3 \right)^{1/2}$.

We can now use equation (\ref{eq:Lnu_fastcool}) to rescale the repeater's observed quiescent flux density to other sources. We assume a power-law energy injection rate to the nebula using similar notation as for the detection-metric defined in the previous section, $\dot{E}_{\rm n} = \tilde{L} t^{-\alpha}$, and that the nebula size is set by the outer ejecta radius so that $R_{\rm n} = v_{\rm ej} t$. This implies that the predicted flux at frequency $\nu$ and time $t$ is
\begin{align} \label{eq:Fnu_quiescent}
F_\nu(t) \lesssim F_{\nu_{\rm r}} &\left(\frac{D}{D_{\rm r}}\right)^{-2} \left(\frac{\nu}{\nu_{\rm r}}\right)^{-p/2} 
\left(\frac{\tilde{L}}{\tilde{L}_{\rm r}}\right)^{(p+2)/4} 
\\ \nonumber
&\times 
\left(\frac{v_{\rm ej} t_{\rm r}}{R_{\rm VLBI}}\right)^{3(2-p)/4}
\left(\frac{t}{t_{\rm r}}\right)^{1 - p/2 - \alpha(p+2)/4} 
\\ \nonumber
&\underset{\alpha=0}{\approx} 188 \, \mu{\rm Jy} \, D_{\rm Gpc}^{-2} \nu_{10}^{-1.15} v_{{\rm ej},9}^{-0.225} 
t_{30\,{\rm yr}}^{-0.15} t_{{\rm r}, 30\,{\rm yr}}^{-0.075}
,
\end{align}
where quantities with subscript $X_{\rm r}$ refer to the assumed/measured properties of the repeater, and the inequality results from the VLBI constraints on the marginally resolved emitting region, $v_{\rm ej, r} t_{\rm r} \lesssim R_{\rm VLBI} \simeq 0.7 \, {\rm pc}$ \citep{Marcote+17} and assuming $p>2$ (otherwise the inequality in equation~\ref{eq:Fnu_quiescent} is reversed).
The last equality in equation~(\ref{eq:Fnu_quiescent}) adopts an electron power-law index of $p=2.3$ consistent with the FRB~121102 spectrum above $10 \, {\rm GHz}$ assuming this is above the cooling frequency.

\begin{figure}
\centering
\includegraphics[width=0.45\textwidth]{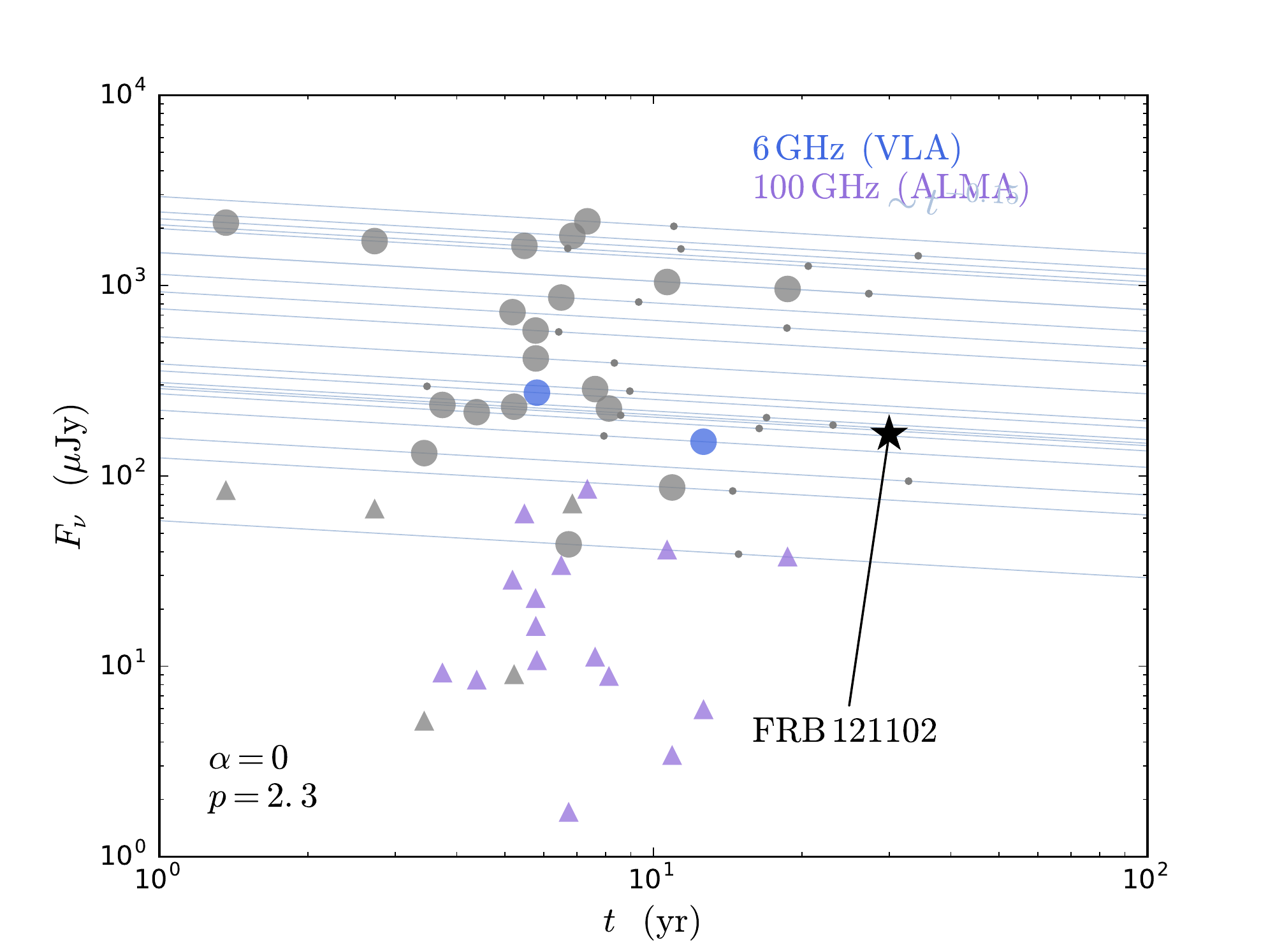}
\includegraphics[width=0.45\textwidth]{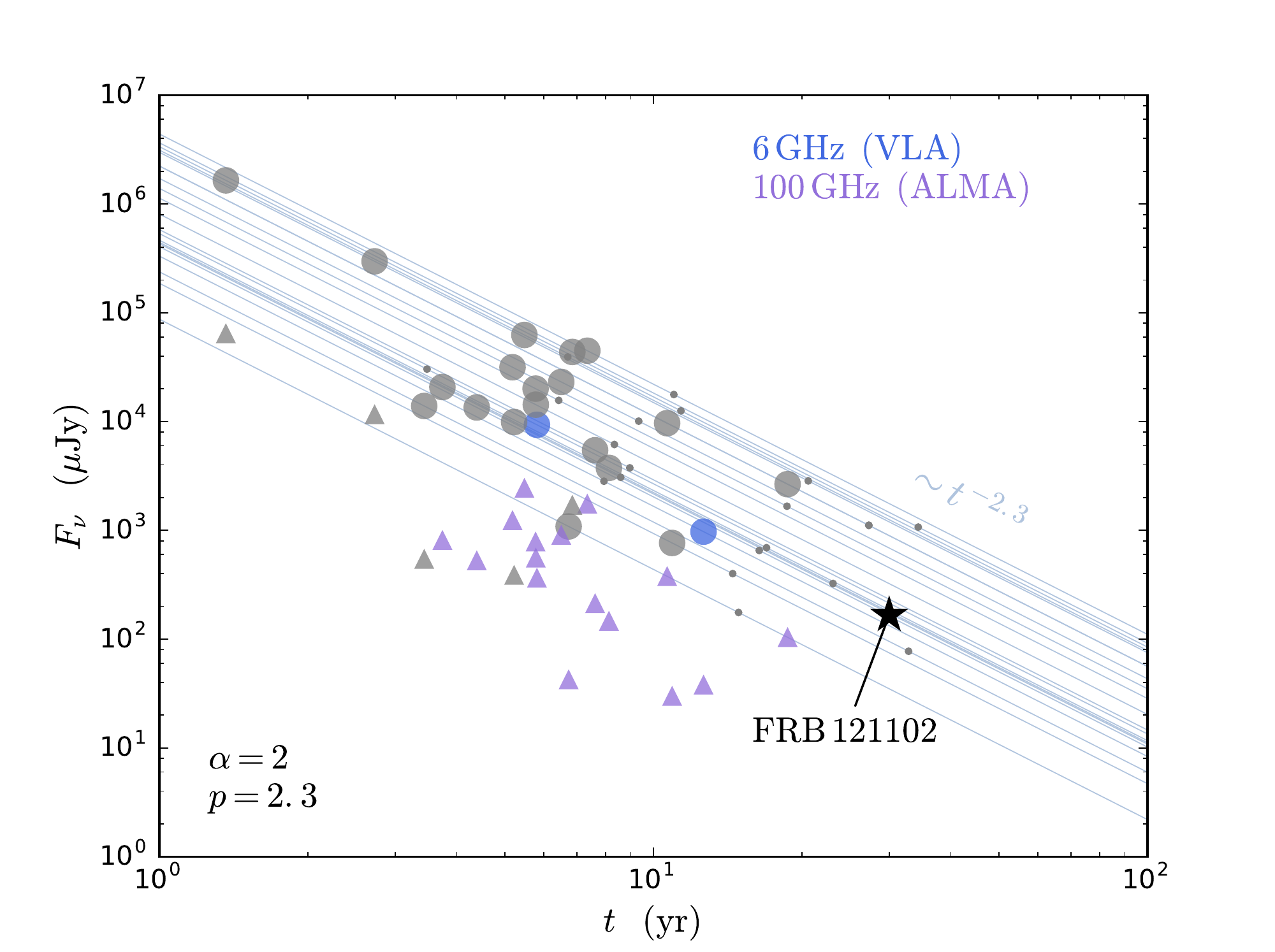}
\caption{Quiescent radio emission predicted for the sample of SLSNe as a function of time at $6 \, {\rm GHz}$ (grey curves) and at the current observed age of each SLSN (blue circles; purple triangles illustrate similar predicted radio fluxes at $100 {\rm GHz}$). 
Symbols colored in grey indicate that the SLSN ejecta is not yet transparent to free-free absorption at the given epoch and frequency band such that the predicted signal implied by equation~\ref{eq:Fnu_quiescent} will not escape (grey dots along the predicted light-curves mark the free-free transparency timescale at $6 \, {\rm GHz}$).
The calculation assumes a naive scaling of the repeater's observed properties, following equation~(\ref{eq:Fnu_quiescent}).
The top panel shows results for a constant nebula energy injection rate ($\alpha=0$), while the bottom panel is for a decaying injection rate proportional to $t^{-2}$, as appropriate for a spin-down powered model ($\alpha=2$).} \label{fig:quiescent}
\end{figure}

Figure~\ref{fig:quiescent} shows the predicted quiescent radio flux in the VLA and ALMA bands assuming an age (time since SN) of $t_{\rm r} = 30 \, {\rm yr}$ for the repeater (see previous section), and for two energy injection models --- $\alpha = 0$ ($2$) heuristically corresponding to a magnetic-dissipation (spin-down) powered models, respectively.
As in equation~(\ref{eq:Fnu_quiescent}), the electron index $p$ is set by the observed spectral slope of the FRB121102 quiescent source, which can be approximate as $F_\nu \sim \nu^{-1.15}$ above $\sim 10 \, {\rm GHz}$, implying $p = 2.3$.

A major underlying assumption in this simplified phenomenological model is that we would observe putative SLSNe quiescent sources within the same spectral region (in this case --- the fast-cooling optically thin regime). This assumption is motivated by the interpretation of the spectral turn-over at $\sim 10 \, {\rm GHz}$ as the cooling break. If this interpretation is correct, then the fact that FRB~121102 is likely observed later after SN then current SLSNe (see previous section) implies that for such SLSNe, the cooling break should be at even lower frequencies, thus validating the implicit assumption of $\nu_{\rm c} < \nu$.

Another assumption of this calculation is that the injected electron spectrum is a power-law.  As already mentioned, if the nebula is powered by the escape of magnetic energy from the engine in a baryon-loaded wind, then the electrons are heated at the wind termination shock to a {\it thermal} energy $k T \sim \bar{\gamma_e} m_e c^{2}$ with a mean Lorentz factor $\bar{\gamma_e} \sim 100$ (equation~\ref{eq:gammabar}) possibly sufficient to explain the observed GHz radio emission of FRB~121102.  In this case, if the baryon loading of the wind is fixed $\xi$, then at earlier times when the magnetic field is higher then the Lorentz factors of electrons contributing in the GHz range will be a part of the Rayleigh-Jeans tail and we will expect a $F_{\nu} \propto \nu^{1/3}$ spectrum \citep[e.g.][]{Giannios&Spitkovsky09}.
The luminosity should scale with magnetic field in this case, $F_\nu \propto B$ such that
\begin{align}
F_\nu &\lesssim F_{\nu_{\rm r}} \left(\frac{D}{D_{\rm r}}\right)^{-2} \left(\frac{\nu}{\nu_{\rm r}}\right)^{1/3} 
\left(\frac{\tilde{L}}{\tilde{L}_{\rm r}}\right)^{1/2} 
\left(\frac{v_{\rm ej} t_{\rm r}}{R_{\rm VLBI}}\right)^{-3/2}
\left(\frac{t}{t_{\rm r}}\right)^{-(2+\alpha)/2} 
\nonumber \\
&\underset{\alpha=0}{\approx} 240 \, \mu{\rm Jy} \, D_{\rm Gpc}^{-2} \nu_{10}^{1/3} v_{{\rm ej},9}^{-3/2} 
t_{30\,{\rm yr}}^{-1} t_{{\rm r}, 30\,{\rm yr}}^{-0.5} .
\end{align}





\section{Conclusions} \label{sec:conclusions}

We have examined the photo-ionization of homologously expanding ejecta by a central ionizing radiation source, with application to GRBs, Type I SLSNe, NS mergers (specifically GW170817), FRBs (focusing on the repeating source FRB~121102), and the very luminous transient ASASSN-15lh.
These diverse phenomena share a commonality --- the possibility that their driving power source is a newly-born magnetar, or otherwise similarly-acting central engine like an accreting black hole.  

Our investigation of the time-dependent ionization state of the expanding ejecta cloud surrounding such a putative central engine is used to address a multitude of its potential observable signatures.  We additionally address the question of whether FRB~121102 is consistent with a young `SLSN-type' magnetar origin, as suggested by e.g. \cite{Metzger+17}, and provide simple analytic models for its observed DM, RM and quiescent radio emission.

Our main conclusions are summarized as follows:
\begin{enumerate}
\item The (density-averaged) ionization fraction of metal rich (e.g. O-rich, pure-Fe) ejecta remains roughly constant in time for an ionizing luminosity source declining as $L_{\rm e} \propto t^{-2}$, as would apply to the late-time magnetar spin-down power.

\item X-rays from SLSNe engines are severely attenuated in the first $\sim$decades post explosion and escape the ejecta due to expansion-dilution rather than classical X-ray break-out \citep{Metzger+14}. This is consistent with X-ray non-detections for the majority of SLSNe and indicates that, except in possible extreme cases, or if density inhomogeneities play an important role, X-rays may not provide the easiest means of testing the magnetar hypothesis for SLSNe.

\item The observed X-ray flux of ASASSN-15lh can only be explained as unabsorbed flux from a central engine if the ejecta mass is assumed to be low ($\lesssim 1 M_\odot$ or $\lesssim 3 M_\odot$ for O-rich or solar composition, respectively).  This is in possible tension with the peak timescale of this event, if this timescale is attributed to the photon-diffusion timescale through the ejecta (though in a TDE, the light curve peak could be set by other effects like the fallback time of the stellar debris).  Alternatively, the X-ray source may be unrelated to the optical/UV transient.

\item For canonical parameters, photo-ionization of SLSNe ejecta can induce significantly larger DM on $\lesssim 10^2 \, {\rm yr}$ timescales than that caused by collisional ionization of shocked matter due to the ejecta-CSM interaction \citep[as advocated by][]{Piro16,Piro&Gaensler18}.  

\item A magnetar central engine operational on $\sim 1-100 \, {\rm d}$ timescales is ruled out for the NS merger GW170817 unless the amount of spin-down power emitted in ionizing radiation is small ($\epsilon_i \ll 1$). Similarly, the hypothesis that the kilonova associated with GW170817 was powered by a central engine (instead of by radioactive decay of freshly synthesized $r$-process material) is ruled out by early X-ray non-detections, unless the ejecta mass is large $\gtrsim 10^{-2} M_\odot$.  However, for such a large ejecta mass, radioactive heating would provide a comparable luminosity to that of the supposed engine, negating the need for the latter.

\item The age of FRB~121102, assuming it originates from a flaring magnetar within a typical Type-I SLSN, is $\gtrsim 30-100 \, {\rm yr}$, consistent with previous analytic estimates \citep{Metzger+17}.

\item The observed DM and upper limits on $\left\vert d{\rm DM}/dt \right\vert$ of FRB~121102 are statistically consistent with the assumption that FRB~121102 originates from a magnetar with properties $\left( B,P_0,M_{\rm ej} \right)$ drawn from the inferred parameters for magnetar-powered Type-I SLSNe \citep{Nicholl+17d}.

\item The high RM of FRB~121102 cannot be caused by the $\gamma \sim 100$ electrons responsible for the associated quiescent radio emission. The RM can be explained by a population of electrons and ions which were injected into the nebula at early times and have since cooled to non-relativistic velocities.

\item Interpreting the observed change in RM as {\it secular} within the framework mentioned above results in an estimate for the repeater's age of $5(\alpha+6) \, {\rm yr}$, where $\dot{E}_{\rm n} \propto t^{-\alpha}$ is the rate of magnetic energy injection to the nebula. This is again consistent with other age constraints on FRB~121102.

\item The maximal number of ejected baryons per unit energy released by a flaring magnetar $\xi_{\rm max}$ (as set by the escape speed from the NS surface) corresponds to a characteristic electron Lorentz factor $\bar{\gamma}_e \sim 10^2$. Remarkably, this value agrees with that required to produce the frequency of the quiescent radio emission coincident with FRB~121102 \citep{Beloborodov17}, providing additional support for the magnetar model.

Future telescopes like UTMOST \citep{Caleb+17} and Apertif \citep{Colegate&Clarke11}, will enable a large expansion in the study of FRB properties,
including those with good localizations,
particularly if all FRBs are accompanied by bright persistent radio sources similar to the quiescent emission of FRB~121102 \citep{Eftekhari+18}.
In the absence of luminous radio nebulae, robust FRB host galaxy association requires higher resolution, sub-arcsecond, localization only accessible to facilities such as VLBA, EVN, VLA, ASKAP, DSA-10 and MeerKAT, which are expensive for this purpose given the large number of observing hours likely required to detect an FRB \citep{Eftekhari&Berger17}.

In addition to expanding the sample size of well-localized repeating FRBs, further monitoring of FRB~121102 may provide crucial information for testing the magnetar hypothesis. In particular, our models predict a secular decline in both DM and RM of the repeater due to the surrounding SN ejecta's expansion. Thus, a falsifiable test of our model, at least in its simplest form, is if both rotation and dispersion measures are not found to decrease (averaging over any random fluctuations) over a baseline of $\sim$several decades.

Finally, we point out the importance of further investigation of X-ray break-out from SLSNe, given that such a signature would provide a smoking-gun indication of a magnetar engine. Though our current analysis suggests that X-rays cannot, for typical `SLSN-type' magnetar and ejecta parameters, ionize their way out of the ejecta, our idealized models assume spherical symmetry and neglect inhomogeneities expected due to e.g. Rayleigh-Taylor instabilities from the nebula-ejecta interface \citep[e.g.][]{Blondin&Chevalier17}. The `fractured' density distribution in this case may allow X-rays to escape at earlier times (and higher luminosities) than predicted by our current spherical models, and we leave investigation of this issue to future work.

We also note that we have focused in this work on SLSNe rather than long-GRB engines because the latter should emit significantly lower luminosity at the late times of interest ($t \gg t_{\rm rot}$). This is a natural consequence of the shorter engine timescale of long-GRBs, $\sim 100 \, {\rm s}$, compared to $\sim {\rm days}$ for SLSNe and the fact that $L \propto \left( t / t_{\rm rot} \right)^{-2}$ for magnetar spin-down \citep[e.g.][]{Margalit+17}. Long-GRB engines are therefore expected to have little effect on the ionization state of their surrounding ejecta on timescales of years or later.

\end{enumerate}

\section*{Acknowledgements}

BM and BDM acknowledge support from NSF award AST-1615084, as well as NASA through the Astrophysics Theory Program grants NNX17AK43G and NNX16AB30G.







\appendix

\section{Temperature Profile of Hydrogen-Rich Ejecta} \label{sec:Appendix_Te}

For hydrogen-rich ejecta, the radial profile of the electron temperature, $T_{\rm e}(r,t)$, can be estimated analytically by considering what sources of heating $\Gamma$ and cooling $\Lambda$ balance on different radial scales.

Absent internal sources of heating (e.g. radioactivity) and neglecting the reverse shock (Appendix \ref{sec:Appendix_reverseshock}), the ejecta heating is determined by the incident radiation field from the central engine.  Compton heating due to inelastic electron scattering occurs at a rate (per unit volume) given by
\begin{align}
\Gamma_{\rm comp} &= f_{\rm ion} n \int \frac{\sigma_{\rm T} u_\nu}{m_e c} h\nu \, d\nu
\\ \nonumber
&\approx f_{\rm ion} n \frac{ \sigma_{\rm T} \left. \nu L_{\rm e,\nu} \right|_{\nu_{\rm min}}}{4\pi m_e c^2 r^2} h\left(\nu_{\rm max}-\nu_{\rm min}\right)
\end{align}
where $\sigma_{\rm T}$ the Thomson cross-section and $u_\nu(r)$ is the radiation energy density.  We have assumed in the second line a logarithmically flat ionizing spectrum, $u_\nu \propto \nu^{-1}$ between $\nu_{\rm min}$ and $\nu_{\rm max}$,
and have neglected radial attenuation of the radiation energy density; the latter is a reasonable approximation for the pure-hydrogen nebula since only photons near the ionization threshold $h\nu \sim 13.6 \, {\rm eV}$ are absorbed.  

Photo-ionization (photo-electric) heating occurs at the rate
\begin{equation}
\Gamma_{\rm pe} = \left( 1 - f_{\rm ion} \right) n \int \frac{ \sigma_{\rm pe}(\nu)c u_\nu}{h \nu} \left( h\nu - h\nu_0 \right) \, d\nu
\approx \alpha_{\rm B} f_{\rm ion}^2 n^2 \frac{h\nu_0}{3}
\end{equation}
where $n$ is the ejecta number density.  In the second equality we have assumed ionization-recombination equilibrium, where $\alpha_{B}$ is the case-B recombination coefficient and $\approx h\nu_0 / 3$ is the mean energy per photo-ionization for a typical dependence $\sigma_{\rm pe}(\nu) \propto \nu^{-3}$ of the cross-section for $\nu \ge \nu_0$ and $u_\nu \propto \nu^{-1}$.

Compton and radiative-recombination cooling can be expressed similarly to the heating terms above,
\begin{align}
\Lambda_{\rm comp} &= f_{\rm ion} n \int \frac{\sigma_{\rm T} u_\nu}{m_e c} 4 k_B T_{\rm e} \, d\nu
\\ \nonumber
&\approx f_{\rm ion} n \frac{ \sigma_{\rm T} \left. \nu L_{\rm e,\nu} \right|_{\nu_{\rm min}}}{4\pi m_e c^2 r^2} 4 k_B T_{\rm e} \ln \left( \frac{\nu_{\rm max}}{\nu_{\rm min}} \right)
\end{align}
and
\begin{equation}
\Lambda_{\rm rr} = \alpha_{\rm B}(T_{\rm e}) f_{\rm ion}^2 n^2 
\left[ \frac{3}{2} + \left(\frac{\partial \ln \alpha_{\rm B}}{\partial \ln T_{\rm e}}\right) \right] k_B T_{\rm e} ,
\end{equation}
where the term in brackets is the average energy loss per recombination.  Finally, free-free cooling occurs at a rate
\begin{equation}
\Lambda_{\rm ff} = f_{\rm ion}^2 n^2 \lambda_{\rm ff}(T_{\rm e})
\approx f_{\rm ion}^2 n^2 \lambda_{\rm ff,0} T_{\rm e}^{1/2} ,
\end{equation}
where $\lambda_{\rm ff}(T_{\rm e}) \approx \lambda_{\rm ff,0} T_{\rm e}^{1/2}$ for temperatures near $T_{\rm e} \sim 10^4 \, {\rm K}$ and $\lambda_{\rm ff,0} \simeq 1.42 \times 10^{-27}$ in appropriate cgs units.
Since we focus here on pure-hydrogen composition, we do not consider line cooling by metals, even though the latter dominates free-free cooling for O-rich ejecta composition.

Balancing various heating and cooling terms ($\Gamma = \Lambda$), we distinguish three regimes relevant at increasing radii within the ejecta.
At small radii, Compton heating balance Compton cooling ($\Gamma_{\rm comp} = \Lambda_{\rm comp}$), and the electron temperature equals the ``Compton Temperature'' of the radiation field,
\begin{equation}
T_{\rm e} = \frac{h \left( \nu_{\rm max} - \nu_{\rm min} \right)}{4 k_B \ln \left( \nu_{\rm max} / \nu_{\rm min} \right)}
\propto r^{0} t^{0} .
\end{equation}
The radially- and temporally-constant value of $T_{\rm e}$ is just a consequence of our assumption that the shape of the spectral energy distribution of the nebula radiation is fixed.  However, the Compton cooling rate decreases with radius as $r^{-2}$ or steeper, such that at sufficiently large radii $\Lambda_{\rm ff} \gg \Lambda_{\rm comp}$, and the temperature is instead set by the balance $\Gamma_{\rm comp} = \Lambda_{\rm ff}$, giving
\begin{equation}
T_{\rm e} \approx \left[ \frac{h \left(\nu_{\rm max}-\nu_{\rm min}\right) \sigma_{\rm T} \left. \nu L_{\rm e,\nu} \right|_{\nu_{\rm min}}}{4\pi \lambda_0 m_e c^2 f_{\rm ion} n r^2} \right]^2
\propto f_{\rm ion}(r,t)^{-2} r^{-4} t^{2} .
\end{equation}
The temporal and radial scaling here apply for the case where $L_{\rm e,\nu} \propto t^{-2}$ and a radially constant (homologously expanding) density. Note that the temperature in this region drops dramatically, roughly as $r^{-4}$, from the $\sim 10^7 \, {\rm K}$ Compton temperature down to $\sim 10^4 \, {\rm K}$ at which photo-electric heating and both free-free and radiative recombination cooling terms become dominant.
Note also that $\Lambda_{\rm comp} \propto t^{-7} T_{\rm e}(t)$ while $\Lambda_{\rm ff} \propto t^{-6} T_{\rm e}(t)^{1/2}$ at a given radius, so that the transition between the Compton cooled and free-free cooled regions moves to smaller radii as time progresses.

Finally, in the outer layers of the ejecta, photo-electric heating is balanced by both free-free and radiative-recombination cooling, which are comparable to one another for $T_{\rm e} \sim 10^4 \, {\rm K}$. Setting $\Gamma_{\rm pe} = \Lambda_{\rm rr}$ results in
\begin{equation}
T_{\rm e} \approx
\frac{h \nu_0}{3 k_B} \left[ \frac{3}{2} + \left(\frac{\partial \ln \alpha_{\rm B}}{\partial \ln T_{\rm e}}\right) \right]^{-1}
\simeq 8.2 \times 10^4 \, {\rm K}
\propto r^0 t^0 ,
\end{equation}
while equating $\Gamma_{\rm pe} = \Lambda_{\rm ff}$ results in
\begin{equation}
T_{\rm e} \sim \left( \frac{h \nu_0 / 3}{\lambda_0 / \alpha_{{\rm B},0}} \right)^{\left[ \frac{1}{2} - \left(\frac{\partial \ln \alpha_{\rm B}}{\partial \ln T_{\rm e}}\right) \right]^{-1}}
\approx 5.3 \times 10^{4} \, {\rm K} 
\propto r^0 t^0 ,
\end{equation}
where we have here written $\alpha_{\rm B}(T_{\rm e}) = \alpha_{{\rm B},0} T_{\rm e}^{\left(\partial \ln \alpha_{\rm B}/\partial \ln T_{\rm e} \right)}$, and assumed for Hydrogen recombination that $\alpha_{{\rm B},0} = 4.68 \times 10^{-10}$, and $\left(\partial \ln \alpha_{\rm B}/\partial \ln T_{\rm e} \right) = -0.8163 - 0.0208 \ln (T_{\rm e}/10^{4}{\rm K})$ in the vicinity of $T_{\rm e} \sim 10^4 \, {\rm K}$ (\citealt{Draine11}; his equation~14.6).

\section{Contribution of the Reverse Shock to Ejecta DM} \label{sec:Appendix_reverseshock}
Here we present a detailed analytic estimate of the maximum DM contributed by the ejecta which has been shock-heated by its interaction with the ambient circumstellar material of assumed density $\rho_{\rm csm}$.  We focus on the early ``ejecta-dominated'' phase, relevant at times $t \ll t_{\rm ST} \sim 10^{3} \, {\rm yr}$ (equation~\ref{eq:t_ST}).  We utilize the solutions described by \cite{Truelove&McKee99} and adopt the same notation as in that paper, to which we refer the reader for additional details on the dynamics of the blast wave and reverse shocks.

The forward shock radius $R_{\rm b}(t)$, for an ejecta of mass $M_{\rm ej}$, total energy $E$, and a density profile characterized by a constant density core and an outer power-law envelope $\rho \propto v^{-n}$, is given by the following implicit relationship,
\begin{equation} \label{eq:Appendix_Rb}
\frac{R_{\rm b}^*}{t^*} = \left(\frac{\alpha}{2}\right)^{-1/2} \ell_{\rm ED} \left[ 1 + \frac{n-3}{3} \left(\frac{\phi_{\rm ED}}{\ell_{\rm ED} f_n}\right)^{1/2} {R_{\rm b}^*}^{3/2} \right]^{-2/(n-3)} .
\end{equation}
Here the dimensionless physical variables demarcated $X^* \equiv X/X_{\rm ch}$ are normalized by their characteristic values,
\be M_{\rm ch} = M_{\rm ej}; \,\,\,R_{\rm ch} = M_{\rm ej}^{1/3} \rho_{\rm csm}^{-1/3}; \,\,\ t_{\rm ch} = E^{-1/2} M_{\rm ej}^{5/6} \rho_{\rm csm}^{-1/3},
\ee
where $\alpha$, $\phi_{\rm ED}$ and $\ell_{\rm ED}$ are constants which depend on the power-law index $n$ \citep{Truelove&McKee99}.  The reverse shock radius $R_{\rm r}$ in this ejecta-dominated phase is simply related to the blast-wave radius by the lead factor, i.e.~$R_{\rm r} = R_{\rm b} / \ell_{\rm ED}$.

At times $t^* \ll t_{\rm CN}^*$ the first term in brackets in equation~(\ref{eq:Appendix_Rb}) is the dominant one, and the solution reduces to free expansion,
\begin{equation}
R_{\rm b}^* (t \ll t_{\rm CN}^*) \approx \left(\frac{\alpha}{2}\right)^{-1/2} \ell_{\rm ED} t^* .
\end{equation}
where
\begin{align}
t_{\rm CN}^* &= \left(\frac{n-3}{3}\right)^{-2/3} \left(\frac{\phi_{\rm ED}}{f_n}\right)^{-1/3} \ell_{\rm ED}^{-2/3} \left(\frac{\alpha}{2}\right)^{1/2}\underset{n=6}{\simeq} \nonumber \\
 &6.47 \times 10^{-5} \left(\frac{w_{\rm core}}{10^{-2}}\right)^2 
\end{align}
is the onset time of the \citet{Chevalier82}, \citet{Nadezhin85} solution.  Here $f_{\rm n}$ is another constant defined by \citet{Truelove&McKee99}, $w_{\rm core} \equiv v_{\rm core} / v_{\rm ej}$, and $v_{\rm core}$ is the velocity at which the density transitions between the flat core and power-law envelope.  Note that the second equality above is given in the limit $w_{\rm core} \ll 1$.

At late times $t^* \gtrsim t_{\rm CN}^*$, the second term in brackets of (\ref{eq:Appendix_Rb}) instead dominates and the blast wave radius instead evolves as a power-law that depends on the density profile,
\begin{align} \label{eq:Appendix_Rb_2}
R_{\rm b}^* (t_{\rm CN}^* \lesssim t^* \lesssim t_{\rm core}^*) &\approx
\left(\frac{n-3}{n}\right)^{-2/n} \left(\frac{\alpha}{2}\right)^{-(n-3)/2n} \ell_{\rm ED}^{(n-2)/n} 
\\ \nonumber
&\times \left(\frac{\phi_{\rm ED}}{f_n}\right)^{-1/n} {t^*}^{(n-3)/n} .
\end{align}
This persists until the time $t_{\rm core}^*$ at which the reverse shock reaches the core-envelope transition, which we estimate from equation~(\ref{eq:Appendix_Rb_2}) to be
\begin{equation}
t_{\rm core}^* \approx \left(\frac{n-3}{3}\right)^{-2/3} \phi_{\rm ED}^{-1/3} \ell_{\rm ED}^{-2/3} f_n^{1/3} \left(\frac{\alpha}{2}\right)^{1/2} w_{\rm core}^{-n/3}
\underset{n=6}{\simeq} 0.647,
\end{equation}
independent of $w_{\rm core}$.

Given these expressions for the shock dynamics, we now estimate the accumulation of shocked ejecta with time. Using expressions for the ejecta mass above normalized velocity coordinate $w_{\rm r} = R_{\rm r}^* / v_{\rm ej}^* t^*$ and remembering that $R_{\rm r}^*(t^*) = R_{\rm b}^*(t^*) / \ell_{\rm ED}$ in the ejecta-dominated phase applicable at $t^* \lesssim t_{\rm core}^*$, we find
\begin{equation}
M_{\rm sh, r}^* (t^*) = \frac{1 - w_{\rm r} (t^*)^{-(n-3)}}{1 - (n/3) w_{\rm core}^{-(n-3)}} .
\end{equation}
Expanding equation~(\ref{eq:Appendix_Rb}) as a Taylor-series in $1-w_{\rm r} \ll 1$ (as applicable at $t^* \ll t_{\rm CN}^*$), and using the approximate free-expansion solution, we find
\begin{equation}
1 - w_{\rm r} (t^*) \approx \frac{2}{3} \left(\frac{\phi_{\rm ED}}{f_n}\right)^{1/2} \left(\frac{\alpha}{2}\right)^{-3/4} \ell_{\rm ED} {t^*}^{3/2} ,
\end{equation}
and thus $M_{\rm sh, r}^* (t^* \ll t_{\rm CN}^*) \propto {t^*}^{3/2}$.
At later times, $t_{\rm CN}^* \lesssim t^* \lesssim t_{\rm core}^*$, it is easy to show that $w_{\rm r} \propto {t^*}^{-3/n}$, and thus $M_{\rm sh, r}^* (t_{\rm CN}^* \lesssim t^* \lesssim t_{\rm core}^*) \propto {t^*}^{3(n-3)/n}$.

The ionization fraction of the ejecta, as results from heating due to the reverse shock, depends on details such as the ejecta composition and cooling, both radiative and from subsequent adiabatic expansion. Here we estimate the {\it largest} possible contribution to the ionized ejecta by making the generous assumption of negligible cooling and complete ionization of shocked matter at all subequent times.  In this case, the mass-averaged ionization fraction of the ejecta simply becomes
\begin{equation}
\left\langle f_{\rm ion} \right\rangle_m = M_{\rm sh, r}^* (t^*) .
\end{equation}
The density-averaged ionization fraction, relevant to calculating the DM, depends on the post-shock density profile, which is not easily described analytically. However, the distribution of shocked ejecta matter between $R_{\rm r}$ and the contact discontinuity will introduce at most an order unity correction to $\left\langle f_{\rm ion} \right\rangle_\rho$.  Again taking the most conservative scenario (largest possible $\left\langle f_{\rm ion} \right\rangle_\rho$) in which the entire shocked mass is concentrated at the reverse shock, $\rho \sim \delta \left( r - R_{\rm r} \right)$, we find
\begin{align}
\left\langle f_{\rm ion} \right\rangle_\rho &\lesssim \frac{n-1}{3n} \left(\frac{\alpha}{2}\right)^{-1} \ell_{\rm ED}^2 w_{\rm core}^2 M_{\rm sh, r}^* \left(\frac{{R_{\rm b}^*} }{  {t^*}}\right)^{-2} 
\\ \nonumber
&\propto
\begin{cases}
{t^*}^{-1/2} &, t^* < t_{\rm CN}^*
\\
{t^*}^{(n-3)/n} &, t_{\rm CN}^* < t^* < t_{\rm core}^*
\end{cases}
.
\end{align}
The density-averaged ionization fraction, and thus the DM, initially decreases with time in the free expansion phase, before increasing again at $t^* > t_{\rm CN}^*$ (this result is only applicable to an ejecta with $n \gtrsim 5$). Since the $n > 5$ solution must converge to the $n=0$ solution at $t^* \gtrsim t_{\rm core}^* $, this implies that the ratio between the DM predicted by the $n>5$ ejecta at early times and the constant density ($n=0$) ejecta is limited to a maximum value,
\begin{equation}
\frac{{\rm DM}_{n > 5}}{{\rm DM}_{n=0}} = \frac{\left\langle f_{\rm ion} \right\rangle_{\rho, n>5}}{\left\langle f_{\rm ion} \right\rangle_{\rho, n=0}} \gtrsim \left( \frac{t_{\rm CN}^*}{t_{\rm core}^*} \right)^{(n-3)/n} = w_{\rm core}^{(n-3)/3} . \label{eq:DMratio}
\end{equation}
Thus, the maximum DM discrepancy between envelope-less and $n>5$ envelope ejecta models is related only to the ratio between the core-envelope transition velocity and the outer (fastest) ejecta velocity. Given the total ejecta mass/energy budget, $w_{\rm core}$ can be expressed in terms of the (uncertain) outer ejecta velocity as
\begin{align}
w_{\rm core} &\underset{w_{\rm core} \ll 1}{\approx} \left[ \frac{10(n-5)E}{3(n-3)M_{\rm ej} v_{\rm ej}^2} \right]^{1/2} 
\\ \nonumber
&\underset{n=6}{\simeq} 2.5 \times 10^{-2} \left(\frac{E}{10^{52}\,{\rm ergs}}\right)^{1/2} \left(\frac{M_{\rm ej}}{10 M_\odot}\right)^{-1/2} \left(\frac{v_{\rm ej}}{c}\right)^{-1} .
\end{align}
For the reasonable assumption that $v_{\rm ej} < c$, we find $w_{\rm core} \sim 0.1$ as a reasonable estimate, in which case the ratio (\ref{eq:DMratio}) is at most a factor of a few for reasonable $n$.

Finally, note that the swept-up shocked circumstellar material can also be important in contributing to the ionized column density. In the $n=0$ case, this is only important after $t \gtrsim t_{\rm ST}$, and thus on longer timescales than of typical interest in our scenario. For the case of an $n > 5$ ejecta, the swept-up mass exceeds (yet remains comparable to) the shocked-ejecta mass already following $t \gtrsim t_{\rm CN}$; however the swept-up circumstellar mass only dominates the shocked ejecta $M_{\rm sh,r}$ by a significant amount at very late times $t \gtrsim t_{\rm ST}$.

\section{DM Probability Distributions} \label{sec:Appendix_P_of_DM}

The probability distribution of dispersion measures and their derivatives can be computed using our \cloudy~photoionization calculations and under the assumption that the magnetar/ejecta parameters inferred for the adopted sample of SLSNe is characteristic of the underlying population.
Focusing in particular on the chance of detecting an FRB with given DM and $d{\rm DM}/dt$, we assign an arbitrary probability metric of FRB detectability as a function of time, adopting a power-law parameterization, $\prob (t) \propto t^{-\alpha}$ with a detectable activity lifetime $t_{\rm a}$. We expect $\alpha$ and $t_{\rm a}$ to be related to some physical measure of burst detectability such as a possible decay in burst luminosity or repetition frequency with time, and adopt toy models with $\alpha =0$ and $\alpha = 2$ and various values for $t_{\rm a}$.

The probability density function (PDF) of detecting an FRB with dispersion measure DM, given that the assumed SLSN progenitor has a free-free transparency time $t_{\rm ff}$ at radio frequencies and dispersion measure ${\rm DM}_{\rm ff}$ at this time, is then
\begin{equation} \label{eq:P(DM)_relationto_P(t)}
\prob \left( \DM \vert \tff, \DMff \right) = 
\prob \left( t \left[\DM ; \tff,\DMff \right] \vert \tff \right) \times \left\vert \frac{d\DM}{dt}\right\vert^{-1} ,
\end{equation}
where $t(\DM)$ is given by inverting the dispersion measure temporal behavior. For O-rich ejecta we have shown in \S~\ref{sec:DM_freefree} that the ionization fraction of the ejecta remains approximately constant, and therefore $\DM(t) = \DMff \left({t}/{\tff}\right)^{-2}$.
Using this, in conjunction with the detection PDF at time $t$
\begin{equation} \label{eq:P(t)}
\prob \left( t \vert \tff \right) = 
\frac{1-\alpha}{\ta^{1-\alpha}-\tff^{1-\alpha}}
\begin{cases}
t^{-\alpha} , &\tff < t < \ta
\\
0 , &{\rm else}
\end{cases}
\end{equation}
normalized such that the integrated distribution over the detectable time-slot $\tff < t < \ta$ is unity.

Combining the above equations we arrive at an expression for the PDF of measuring an FRB with dispersion measure DM, {\it for a given set of SLSN parameters $\tff$, $\DMff$}. It is
\begin{equation}
\prob \left( \DM \vert \tff,\DMff \right) = \frac{1-\alpha}{2 \DMff} \left[ \left(\frac{\ta}{\tff}\right)^{1-\alpha} - 1 \right]^{-1} \left(\frac{\DM}{\DMff}\right)^{(\alpha-3)/2} .
\end{equation}

To complete the calculation, we use the {\it distribution} of SLSNe parameters $\tff$, $\DMff$ as found by our \cloudy~calculations and apply Bayes' law to obtain
\begin{equation}
\prob \left(\DM\right) = \int d\tff d\DMff \, \prob\left(\DM \vert \tff,\DMff \right) \prob\left(\tff,\DMff\right) .
\end{equation}
where, for our finite sample of SLSNe parameters
\begin{equation}
\prob \left( \tff, \DMff \right) \approx \frac{1}{N} \sum_{i=1}^{N} \delta(\tffi) \delta(\DMffi) ,
\end{equation}
and thus, the final observable FRB DM distribution is
\begin{align}
\prob(\DM) \approx &\frac{1}{N} \sum_{i=1}^{N} 
\frac{1-\alpha}{2 \DMffi} \left[ \left(\frac{\ta}{\tffi}\right)^{1-\alpha} - 1 \right]^{-1} \left(\frac{\DM}{\DMffi}\right)^{(\alpha-3)/2}
\nonumber \\
&\times
\begin{cases}
1 , &\frac{\DM}{\DMffi} \in \left[ 1, \left(\frac{\ta}{\tffi}\right)^{-2} \right]
\\
0 , &{\rm else}
\end{cases}
\end{align}
where the index $i$ enumerates the free-free transparency time $\tffi$ and the dispersion measure at that time $\DMffi$ for the $i$th SLSN in our sample.

Since ${d\DM}/{dt} = -2 \DM / t$, a similar analysis can be performed for the DM derivative, resulting in
\begin{align}
\prob\left(\frac{d\DM}{dt}\right) &\approx \frac{1}{N} \sum_{i=1}^{N} 
\frac{(1-\alpha) \tffi}{2 \DMffi} \left[ \left(\frac{\ta}{\tffi}\right)^{1-\alpha} - 1 \right]^{-1} \left(-\frac{\tffi}{2\DMffi}\frac{d\DM}{dt}\right)^{(\alpha-4)/3}
\\ \nonumber
&\times
\begin{cases}
1 , &\left\vert \frac{d\DM}{dt} \right\vert \in \frac{2\DMffi}{\tffi} \left[ 1, \left(\frac{\ta}{\tffi}\right)^{-3} \right]
\\
0 , &{\rm else}
\end{cases}
\end{align}

\bibliographystyle{yahapj}
\bibliography{refs}


\bsp	
\label{lastpage}
\end{document}